\documentclass[MSNbibl,nameyear,seceqn,dvips]{arxstspdf}
\usepackage{dcolumn,multirow}
\usepackage[safe,T1]{tipa}     %%% uzsikrauti
\usepackage{graphicx}
\usepackage{flushend}
\usepackage{stfloats}

\volume{29}
\issue{3}
\pubyear{2014}
\firstpage{323}
\lastpage{358}
\doi{10.1214/14-STS480} % kopijuoti is 'New paper accepted'
\docsubty{FLA}

\makeatletter
\def\polhk#1{\textpolhook{#1}} %%% prie apibrezimu
\newcolumntype{d}[1]{D{.}{.}{#1}}
\newcommand{\comp}{\mbox{complier}}
\newcommand{\never}{\mbox{never-taker}}
\newcommand{\always}{\mbox{always-taker}}
\newcommand{\defier}{\mbox{defier}}
\newcommand{\ifff}{\mbox{if }}
\newcommand{\iv}{\mathrm{iv}}
\newcommand{\tsls}{\mathrm{tsls}}
\newcommand{\ils}{\mathrm{ils}}
\newcommand{\liml}{\mathrm{liml}}
\newcommand{\pr}{\operatorname{pr}}
\newcommand{\obs}{{\mathrm{obs}}}
\newcommand{\oy}{\overline{Y}}
\newcommand{\ox}{\overline{X}}
\newcommand{\oz}{\overline{Z}}
\newcommand{\mme}{{\mathbb{E}}}
\makeatother

\begin{document}
\begin{frontmatter}

\title{Instrumental Variables: An~Econometrician's Perspective\thanksref{T1}}
% kai straipsnis turi susijusiu diskusiju ir rejoinder'iu
\relateddois{T1}{Discussed in \relateddoi{d}{10.1214/14-STS494},
\relateddoi{d}{10.1214/14-STS485}, \relateddoi{d}{10.1214/14-STS488},
\relateddoi{d}{10.1214/14-STS491}; rejoinder at \relateddoi{r}{10.1214/14-STS496}.}
\runtitle{Instrumental Variables}

\begin{aug}
\author[a]{\fnms{Guido W.}~\snm{Imbens}\corref{}\ead[label=e1]{imbens@stanford.edu}%
\ead[label=u1,url]{http://www.gsb.stanford.edu/users/imbens}}
\runauthor{G. W. Imbens}

\affiliation{Stanford University}

\address[a]{Guido W. Imbens is the Applied Econometrics Professor and Professor of Economics,
Graduate School of Business, Stanford University, Stanford, California 94305, USA and
NBER \printead{e1,u1}.}
\end{aug}

% ABSTRACT
%
\begin{abstract}
I review recent work in the statistics literature on instrumental
variables methods from an econometrics perspective. I discuss some of
the older, economic, applications including supply and demand models
and relate them to the recent applications in settings of randomized
experiments with noncompliance. I discuss the assumptions underlying
instrumental variables methods and in what settings these may be
plausible. By providing context to the current applications, a better
understanding of the applicability of these methods may arise.
\end{abstract}

% KEYWORDS
\end{frontmatter}

\setcounter{footnote}{1}

%s1 #&#
\section{Introduction}

Instrumental Variables (IV) refers to a set of methods developed in
econometrics starting in the 1920s to draw causal inferences in
settings where the treatment of interest cannot be credibly viewed as
randomly assigned, even after conditioning on additional covariates,
that is, settings where the assumption of no unmeasured confounders
does not hold.\footnote{There is another literature in econometrics
using instrumental variables methods also to deal with classical
measurement error (where explanatory variables are measured with error
that is independent of the true values). My remarks in the current
paper do not directly reflect on the use of instrumental variables to
deal with measurement error. See \citet{Sar58} for a classical paper,
and Hillier (\citeyear{Hil90}) and \citet{Are02} for more recent discussions.}
In the last two decades, these methods have attracted considerable
attention in the statistics literature. Although this recent statistics
literature builds
on the earlier econometric literature, there are nevertheless
important differences.
First, the recent statistics literature
primarily focuses on the binary treatment case. Second, the recent
literature explicitly allows for treatment effect heterogeneity. Third,
the recent instrumental variables literature (starting with \cite*{ImbAng94}; \cite*{AngImbRub}; \cite*{Hec}; \cite*{Man90}; and \cite*{Rob86}) explicitly uses the potential outcome framework
used by Neyman for randomized experiments and generalized to
observational studies by
Rubin (\citeyear{Rub74}, \citeyear{Rub78}, \citeyear{Rub90}).
Fourth, in the
applications this literature has concentrated on, including randomized
experiments with noncompliance, the intention-to-treat or reduced-form
estimates are often of greater interest than they are in the
traditional econometric simultaneous equations applications.

Partly the recent statistics literature has been motivated by the earlier
econometric literature on instrumental variables, starting with \citet{Wri28} (see the discussion on the origins of instrumental variables in
\cite*{StoTre03}). However, there are also other antecedents,
outside of the traditional econometric instrumental variables
literature, notably the work by Zelen on encouragement designs (Zelen, \citeyear{Zel79}, \citeyear{Zel90}).
Early papers in the recent statistics literature include \citet{AngImbRub}, \citet{Rob} and
\citet{McCNew94}.
Recent reviews include \citet{Ros10}, \citet{Vanetal11} and
\citet{HerRob06}. Although these reviews include
many references to the earlier economics literature, it might still be
useful to discuss the econometric literature in more detail to provide
some \mbox{background} and perspective on the applicability of instrumental
variables methods in other fields. In this discussion, I~will do so.

Instrumental variables methods have been a central part of the
econometrics canon since the first half of the twentieth century,
and\vadjust{\goodbreak}
continue to be an integral part of most graduate and undergraduate
textbooks (e.g., Angrist and Pischke, \citeyear{AngPis09}; \cite*{BowTur84}; \cite*{Gre11}; \cite*{Hay00};  \cite*{Man95}; \cite*{StoWat10}; Wooldridge, \citeyear{Woo10}, \citeyear{Woo08}).
Like the statisticians Fisher
and Neyman (\cite*{Fis25}; Splawa-Neyman, \citeyear{Spl90}),
early econometricians such as \citet{Wri28}, \citet{Wor27},
\citet{Tin30} and\break \citet{Haa43} were interested in drawing causal
inferences, in their case about the effect of economic policies on
economic behavior. However, in sharp contrast to the statistical
literature on causal inference,
%in observational studies by researchers such as \citet{Coc68}, Rubin
%(1974, 2006), \reftext{Rosenbaum (2009, \citeyear{Ros10})} who build on the Fisher and
%Neyman tradition,
the starting point for these econometricians was \textit{not} the
randomized experiment. From the outset, there was a recognition that in
the settings they studied, the causes, or treatments, were not {assigned} to passive units (economic agents in their setting, such as
individuals, households, firms or countries). Instead the economic
agents actively influence, or even explicitly choose, the level of the
treatment they receive. Choice, rather than chance, was the starting
point for thinking about the assignment mechanism in the econometrics
literature. In this perspective, units receiving the active treatment
are different from those receiving the control treatment not just
because of the receipt of the treatment: they (choose to) receive the
active treatment because they are different to begin with. This makes
the treatment potentially \textit{endogenous}, and creates what is
sometimes in the econometrics literature referred to as the \textit{selection problem} (\cite{Hec79}).

The early econometrics literature on instrumental variables did not
have much impact
on thinking in the statistics community. Although some of the technical
work on large sample properties of various estimators did get published
in statistics journals (e.g., the still influential Anderson and
Rubin, \citeyear{A49} paper), applications by noneconomists were rare. It is not
clear exactly what the reasons for this are. One possibility is the
fact that the early literature on instrumental variables was closely
tied to substantive economic questions (e.g., interventions in
markets), using theoretical economic concepts that may have appeared
irrelevant or difficult to translate to other fields (e.g.,
supply and demand). This may have suggested to noneconomists that the
instrumental variables methods in general had limited applicability
outside of economics. The use of economic concepts was not entirely
unavoidable, as the critical assumptions underlying instrumental
variables methods are substantive and require subtle subject matter knowledge.
%Often that aspect is lost in the recent statistics literature where
%instrumental variables methods are still viewed with suspicion.
A second reason may be that although the early work by Tinbergen and
Haavelmo used a notation that is very similar to what \citet{Rub74}
later called the potential outcome notation, quickly the literature
settled on a notation only involving realized or observed outcomes; see
for a historial perspective \citet{autokey71} and \citet{Imb97}.
This realized-outcome notation that remains common in the econometric
textbooks obscures the connections between the Fisher and Neyman work
on randomized experiments and the instrumental variables literature. It
is only in the 1990s that econometricians returned to the potential
outcome notation for causal questions (e.g., \cite*{Hec}; \cite*{Man90}; \cite*{ImbAng94}), facilitating and initiating a
dialogue with statisticians on instrumental variable methods.

The main theme of the current paper is that the early work in
econometrics is helpful in understanding the modern instrumental
variables literature, and furthermore, is potentially useful in
improving applications of these methods and identifying potential instruments.
These methods may in fact be useful in many settings statisticians
study. Exposure to treatment is rarely solely a matter of chance or
solely a matter of choice. Both aspects are important and help to
understand when causal inferences are credible and when they are not.
In order to make these points, I will discuss some of the early work
and put it in a modern framework and notation. In doing so, I will
address some of the concerns that have been raised about the
applicability of instrumental variables methods in statistics.
I will also discuss some areas where the recent statistics literature
has extended and improved our understanding of instrumental variables
methods. Finally, I will review some of the econometric terminology and
relate it to the statistical literature to remove some of the semantic
barriers that continue to separate the literatures.
I should emphasize that many of the topics discussed in this review
continue to be active research areas, about which there is considerable
controversy both inside and outside of econometrics. %The views
%presented here are largely mine and should not be taken as a consensus
%view of the econometrics profession. %, although I will also present
%some different viewpoints.

The remainder of the paper is organized as follows.
In Section~\ref{section:choice}, I will discuss the distinction
between the statistics literature on causality with its primary focus
on chance, arising from its origins in the experimental literature, and
the econometrics or economics literature with its emphasis on choice.
The next two sections discuss in detail two classes of examples.
In Section~\ref{section:supplydemand}, I discuss the canonical example
of instrumental variables in economics, the estimation of supply and
demand functions.
In Section~\ref{section:modern}, I discuss a modern class of examples,
randomized experiments with noncompliance.
In
Section~\ref{section:content}, I discuss the substantive content of
the critical assumptions, and in Section~\ref{section:textbook}, I
link the current literature to the older textbook discussions.
%Section \ref{section:estimation} contains a short review of methods
%for estimation and inference.
In Section~\ref{section:extensions}, I discuss some of the recent
extensions of traditional instrumental variables methods.
Section~\ref{section:conclusion} concludes.

%s2 #&#
\section{Choice versus Chance in Treatment~Assignment}
\label{section:choice}

Although the objectives of causal analyses in statistics and
econometrics are very similar, traditionally statisticians and
economists have approached these questions very differently.
A
key difference in the approaches taken in the statistical and
econometric literatures is the focus on different assignment
mechanisms, those with an emphasis on chance versus those with an
emphasis on choice.
Although in practice in many observational studies assignment
mechanisms have elements of both chance and choice, the traditional
starting points in the two literatures are very different, and it is
only recently that these literatures have discovered how much they have
in common.\footnote{In both literatures, it is typically assumed that
there is no interference between units. In the statistics literature,
this is often referred to as the \textit{Stable Unit Treatment Value
Assumption} (SUTVA, \cite*{Rub78}). In economics, there are many cases
where this is not a reasonable assumption because there are \textit{general equilibrium} effects. In an interesting recent experiment,
\citet{Creetal}
varied the scale of experimental interventions (job training programs
in their case) in different labor markets and found that the scale
substantially affected the average effects of the interventions. There
is also a growing literature on settings directly modeling interactions.
In this discussion, I will largely ignore the complications arising
from interference between units. See, for example, Manski (\citeyear{Man00N1}).}

%s2.1 #&#
\subsection{The Statistics Literature: The Focus on Chance}

The starting point in the statistics literature, going back to \citet{Fis25} and
Splawa-Neyman (\citeyear{Spl90}), is the randomized experiment, with both
Fisher and Neyman motivated by agricultural applications where the
units of analysis are plots of land. To be specific, suppose we are
interested in the average causal effect of a binary treatment or
intervention, say fertilizer $A$ or fertilizer $B$, on plot yields. In the
modern notation and language originating with \citet{Rub74}, the unit
(plot) level causal effect is a comparison
between the two potential outcomes, $Y_i(A)$ and $Y_i(B)$ [e.g.,
the difference
$\tau_i=Y_i(B)-Y_i(A)$], where $Y_i(A)$ is the potential outcome given
fertilizer $A$ and $Y_i(B)$ is the potential outcome given fertilizer
$B$, both for plot $i$. In a completely randomized experiment with $N$
plots, we select $M$ (with $M\in\{1,\ldots,N-1\}$) plots at random to
receive fertilizer $B$, with the remaining $N-M$ plots assigned to
fertilizer $A$.
Thus, the
treatment assignment,
denoted by $X_i\in\{A,B\}$ for plot $i$, is by design independent of
the potential outcomes.\footnote{To facilitate comparisons with the
econometrics literature, I will follow the notation that is common in
econometrics, denoting the endogenous regressors, here the treatment of
interest, by $X_i$, and later the instruments by $Z_i$. Additional
(exogenous) regressors will be denoted by $V_i$. In the statistics
literature, the treatments of interested are often denoted by $W_i$,
the instruments by $Z_i$, with $X_i$ denoting additional regressors or
attributes.}
In this specific setting, the work by Fisher and Neyman shows how one
can draw exact causal inferences. Fisher focused on calculating exact
$p$-values for sharp null hypotheses, typically the null hypothesis of no
effect whatsoever, $Y_i(A)=Y_i(B)$ for all plots. Neyman focused on
developing unbiased estimators for the average treatment effect $\sum_i(Y_i(A)-Y_i(B))/N$ and the variance of those estimators.

The subsequent literature in statistics, much of it associated with the
work by Rubin and coauthors (\cite*{Coc68}; \cite*{CocRub73};
Rubin, \citeyear{Rub74}, \citeyear{Rub90}, \citeyear{Rub06}; Rosenbaum and Rubin, \citeyear{RosRub83};
\cite*{RubTho92}; Rosenbaum, \citeyear{Ros02}, \citeyear{Ros10}; \cite*{Hol86}) has focused on
extending and generalizing the Fisher and Neyman results that were
derived explicitly for randomized experiments to the more general
setting of observational studies. A large part of this literature
focuses on the case where the researcher has additional background
information available about the units in the study. The additional
information is in the form of pretreatment variables or covariates not
affected by the treatment. Let $V_i$ denote these covariates. A key
assumption in this literature is that conditional on these pretreatment
variables the assignment to treatment is independent of the treatment
assignment. Formally,
\[
%begin{equation}\label{eq:unconf}
X_i \perp \bigl(Y_i(A),  Y_i(B)\bigr) \vert
V_i \quad \mbox{(unconfoundedness)}.
\]
%
%end{equation}
Following \citet{Rub90}, I refer to this assumption as \textit{unconfoundedness given} $V_i$, also known as \textit{no unmeasured
confounders}. This assumption, in combination with the auxiliary
assumption that for all values of the covariates the probability of
being assigned to each level of the treatment is strictly positive is
referred to as
\textit{strong ignorability} (Rosenbaum and Rubin, \citeyear{RosRub83}). If we assume
only that $X_i\perp Y_i(A)|V_i$ and $X_i\perp Y_i(B)|V_i$ rather than
jointly, the assumption is referred to as \textit{weak unconfoundedness}
(\cite{Imb00}), and the combination as \textit{weak ignorability}.
Substantively, it is not clear that there are cases in the setting with
binary treatments where the weak version is plausible but not the
strong version, although the difference between the two assumptions has
some content in the multivalued treatment case (\cite{Imb00}).
In the econometric literature, closely related assumptions are referred
to as
\textit{selection-on-observables} (\cite{BarCaiGol80}) or
\textit{exogeneity}.

Under weak ignorability (and thus also under strong ignorability), it
is possible to estimate precisely the average effect of the treatment in large
samples. In other words, the average effect of the treatment is \textit{identified}.
Various specific methods have been proposed, including matching,
subclassification and regression. See \citet{Ros10}, \citet{Rub06},
Imbens (\citeyear{Imb04}, \citeyear{ImbN2}), \citet{GelHil06}, \citet{ImbRub}  and \citet{AngPis09} for general discussions and surveys.
Robins and coauthors (\cite*{Rob86}; \cite*{GilRob01}; \cite*{RicRob}; Van der Laan and Robins, \citeyear{vanRob03}) have extended this
approach to settings with sequential treatments.

%s2.2 #&#
\subsection{The Econometrics Literature: The~Focus~on~Choice}

In contrast to the statistics literature whose point of departure was
the randomized experiment, the starting point in the economics and
econometrics literatures for studying causal effects emphasizes the
choices that led to the treatment received. Unlike the original
applications in statistics where the units are passive, for example,
plots of land, with no influence over their treatment exposure, units
in economic analyses are typically economic agents, for example,
individuals, families, firms or administrations. These are agents with
objectives and the ability to pursue these objectives within constraints.
The objectives are typically closely related to the outcomes under the
various treatments.
The constraints may be legal, financial or information-based.

The starting point of economic science is to model these agents as
behaving optimally. More specifically, this implies that economists
think of everyone of these agents as choosing the level of the
treatment to most efficiently pursue their objectives given the
constraints they face.\footnote{In principle, these objectives may
include the effort it takes to find the optimal strategy, although it
is rare that these costs are taken into account.}
In practice, of course, there is often evidence that not all
agents behave optimally. Nevertheless, the starting point is the
presumption that optimal behavior is a reasonable approximation to
actual behavior, and the models economists take to the data often
reflect this.

%s2.3 #&#
\subsection{Some Examples}

Let us contrast the statistical and econometric approaches in a highly
stylized example.
\citet{Roy51} studies the problem of occupational choice and the
implications for the observed distribution of earnings. He focuses on
an example where individuals can choose between two occupations,
hunting and fishing. Each individual has a level of productivity
associated with each occupation, say, the total value of the catch per day.
For individual $i$, the two productivity levels
are $Y_i(h)$ and $Y_i(f)$, for the productivity level if hunting and
fishing, respectively.\footnote{In this example, the no-interference
(SUTVA) assumption that there are no effects of other individual's
choices and, therefore, that the individual level potential outcomes
are well defined is tenuous---if one hunter is successful that will
reduce the number of animals available to other hunters---but I will
ignore these issues here.}
Suppose the researcher is interested in the average difference in
productivity in these two occupations, $\tau=\mathbb{E}[Y_i(f)-Y_i(h)]$, where the averaging is over the population of
individuals.\footnote{That is not actually the goal of Roy's original
study, but that is beside the point here.}
The researcher observes for all units in the sample the occupation they
chose ($X_i$, equal to $h$ for hunters and $f$ for fishermen) and the
productivity in their chosen occupation,
\[
Y_i^\obs=Y_i(X_i)=
\cases{Y_i(h) & $\mbox{if } X_i=h$,\vspace*{2pt}
\cr
Y_i(f) & $\mbox{if } X_i=f$.}
\]
In the Fisher--Neyman--Rubin statistics tradition, one might start by
estimating $\tau$ by comparing productivity levels by occupation:
\[
\hat{\tau}=\oy^\obs_f-\oy^\obs_h,
\]
where
\begin{eqnarray*}
\oy^\obs_f &=& \frac{1}{N_f}\sum
_{i:X_i=f} Y^\obs_i,\quad
\oy^\obs_h=\frac{1}{N_h}\sum
_{i:X_i=h} Y^\obs_i,
\\
N_f &=& \sum_{i=1}^N
\mathbf{1}_{X_i=f} \quad \mbox{and} \quad N_h=N-N_f.
\end{eqnarray*}
If there is concern that these unadjusted differences are not credible
as estimates of the average causal effect, the next step in this
approach would be to
adjust for observed individual characteristics such as education levels
or family background. This would be justified if individuals can be
thought of as choosing, at least within homogenous groups defined by
covariates, randomly which occupation to engage in.

Roy, in the economics tradition, starts from a very different place.
Instead of assuming that individuals choose their occupation (possibly
after conditioning on covariates) randomly, he assumes that each
individual chooses her occupation optimally, that is, the occupation
that maximizes her productivity:
\[
X_i= \cases{ f & $\mbox{if }Y_i(f)\geq
Y_i(h)$,
\cr
h & $\mbox{otherwise}$.}
\]
There need not be a solution in all cases, especially if there is
interference, and thus there are general equilibrium effects, but I
will assume here that such a solution exists.
If this assumption about the occupation choice were strictly true, it
would be difficult to learn much about $\tau$ from data on occupations
and earnings.
%In the spirit of research by Manski (\citeyear{Man90}, \citeyear{M92}, \citeyear{Man03}, \citeyear{M08}),
%one can
In the spirit of research by Manski (\citeyear{Man90}, \citeyear{Man00N2}, \citeyear{Man01}),
Manski and Pepper (\citeyear{ManPep00}), and Manski et al. (\citeyear{Manetal92N1}),
one can
derive bounds on $\tau$, exploiting the fact that if $X_i=f$,
then the unobserved $Y_i(h)$ must satisfy $Y_i(h)\leq Y_i(f)$, with
$Y_i(f)$ observed. For the Roy model, the specific calculations have
been reported in Manski (\citeyear{Man95}), Section~2.6. Without additional
information or restrictions, these bounds might be fairly wide, and often one
would not learn much about $\tau$. However, the original version of
the Roy model, where individuals know ex ante  the exact value of
the potential outcomes and choose the level of the treatment
corresponding to the maximum of those, is ultimately not plausible in
practice. It is likely that  individuals face uncertainty regarding their future productivity,
and thus may not be able to choose the ex post  optimal
occupation; see for bounds under that scenario \citet{ManNag98}.
Alternatively, and this is emphasized in \citet{AthSte98},
individuals may have more complex objective functions taking into
account heterogenous costs or nonmonetary benefits associated with each
occupation. This creates a wedge between the outcomes that the
researcher focuses on and the outcomes that the agent optimizes over.
What is key here in relation to the statistics literature is that under
the Roy model and its generalizations the very fact that two
individuals have different occupations is seen as indicative that they
have different potential outcomes, thus fundamentally calling into
question the unconfoundedness assumption that individuals with similar
pretreatment variables but different treatment levels are comparable.
This concern about differences between individuals with the same values
for pretreatment variables but different treatment levels underlies
many econometric analyses of causal effects, specifically in the
literature on selection models. See \citet{HecRob} for a
general discussion.

Let me discuss two additional examples. There is a large literature in
economics concerned with estimating the causal effect of educational
achievement (measured as years of education) on earnings; see for
general discussions \citet{Gri77} and \citet{Car01}. One starting
point, and in fact the basis of a large empirical literature, is to
compare earnings for individuals who look similar in terms of
background characteristics, but who differ in terms of educational
achievement. The concern in an equally large literature is that those
individuals who choose to acquire higher levels of education did so
precisely because they expected their returns to additional years of
education to be higher than individuals who choose not to acquire
higher levels of education expected their returns to be. In the
terminology of the returns-to-education literature, the individuals
choosing higher levels of education may have higher levels of ability,
which lead to higher earnings for given levels of education.

Another canonical example is that of voluntary job training programs.
One approach to estimate the causal effect of training programs on
subsequent earnings would be to compare earnings for those
participating in the program with earnings for those who did not. Again
the concern would be that those who choose to participate did so
because they expected bigger benefits (financial or otherwise) from
doing so than individuals who chose not to participate.

These issues also arise in the missing data literature.
%The statistics
%literature (Rubin, \citeyear{Rub76}, \citeyear{}; \cite*{LitRub87})
The statistics literature (Rubin, \citeyear{Rub76}, \citeyear{Rub87}, \citeyear{Rub96}; Little and Rubin, \citeyear{LitRub87})
has primarily
focused on models that assume that units with item nonresponse are
comparable to units with complete response, conditional on covariates
that are always observed. The econometrics literature (Heckman, \citeyear{Hec76},
\citeyear{Hec79}) has focused more heavily on models that interpret the nonresponse
as the \mbox{result} of systematic differences between units.
%Philipson (\citeyear{}, \citeyear{Phi97N2}) takes this
Philipson (\citeyear{Phi97N1}, \citeyear{Phi97N2}), Philipson and DeSimone
(\citeyear{PhiDeS97}), and Philipson and Hedges (\citeyear{PhiHed98}) take this
even further, viewing survey response as
a market transaction, where individuals not responding the survey do so
deliberately because the costs of responding outweighs the benefits to
these nonrespondents.
The Heckman-style selection models often assume strong parametric
alternatives to the Little and Rubin missing-at-random or ignorability
condition. This has often in turn led to estimators that are sensitive
to small changes in the data
%generating process.
generating process. See Little (\citeyear{L85}).

These issues of nonrandom selection are of course not special to
economics. Outside of randomized experiments, the exposure to treatment
is typically also chosen to achieve some objectives, rather than
randomly within homogenous populations. For example, physicians
presumably choose treatments for their patients optimally, given their
knowledge and given other constraints (e.g., financial).
Similarly, in economics and other social sciences one may view
individuals as making optimal decisions, but these are typically made
given incomplete information, leading to errors that may make the
ultimate decisions appear as good as random within homogenous
subpopulations. What is important is that the starting point is
different in the two disciplines, and this has led to the development
of substantially different methods for causal inference.

%s2.4 #&#
\subsection{Instrumental Variables}

How do instrumental variables methods address the type of selection
issues the Roy model raises? At the core, instrumental variables change
the incentives for agents to choose a particular level of the
treatment, without affecting the potential outcomes associated with
these treatment levels. Consider a job training program example where
the researcher is interested in the average effect of the training
program on earnings.
Each individual is characterized by two potential earnings outcomes,
earnings given the training and earnings in the absence of the
training. Each individual chooses to participate or not based on their
perceived net benefits from doing so.
As pointed out in \citet{AthSte98}, it is important that these
net benefits that enter into the individual's decision differ from the
earnings that are the primary outcome of interest to the researcher.
They do so
%for that individual in each regime (participation or not)
by the costs associated with participating in that regime.
Suppose that there is variation in the costs individuals incur with
participation in the training program.
The costs are broadly defined, and may include travel time to the
program facilities, or the effort required to become informed about the program.
Furthermore, suppose that these costs are independent of the potential
outcomes. This is a strong assumption, often made more plausible by
conditioning on covariates. Measures of the participation cost may then
serve as instrument variables and aid in the identification of the
causal effects of the program. Ultimately, we compare earnings for
individuals with low costs of participation in the program with those
for individuals with high costs of participation and attribute the
difference in average earnings to the increased rate of participation
in the program among the two groups.

In almost all cases, the assumption that there is no direct effect of
the change in incentives on the potential outcomes is controversial,
and it needs to be assessed at a case-by-case level. The second part of
the assumption, that the costs are independent of the potential
outcomes, possibly after conditioning on covariates, is qualitatively
very different. In some cases, it is satisfied by design, for example,
if the incentives are randomized. In observational studies, it is a
substantive, unconfoundedness-type, assumption, that may be more
plausible or at least approximately hold after conditioning on
covariates. For example, in a number of studies researchers have used
physical distance to facilities as instruments for exposure to
treatments available at such facilities. Such studies include \citet{McCNew94} and
\citet{Baietal10}
who use distance to hospitals with particular capabilities as an
instrument for treatments associated with those capabilities, after
conditioning on distance to the nearest medical facility, and \citet{Car95}, who uses distance to colleges as an instrument for attending college.

%s3 #&#
\section{The Classic Example: Supply~and~Demand}
\label{section:supplydemand}

In this section, I will discuss the classic example of instrumental
variables methods in econometrics, that is, simultaneous equations.
Simultaneous equations models are both at the core of the econometrics
canon and at the core of the confusion concerning instrumental
variables methods in the statistics literature.
More precisely, in this section I will look at supply and demand models
that motivated the original research into instrumental variables. Here,
the \textit{endogeneity}, that is, the violation of unconfoundedness,
arises from an equilibrium condition.
I will discuss the model in a very specific example to make the issues
clear, as I think that perhaps the level of abstraction used in the
older econometric text books has hampered communication with
researchers in other fields.

%f1
%f1 #&#
\begin{figure*}

\includegraphics{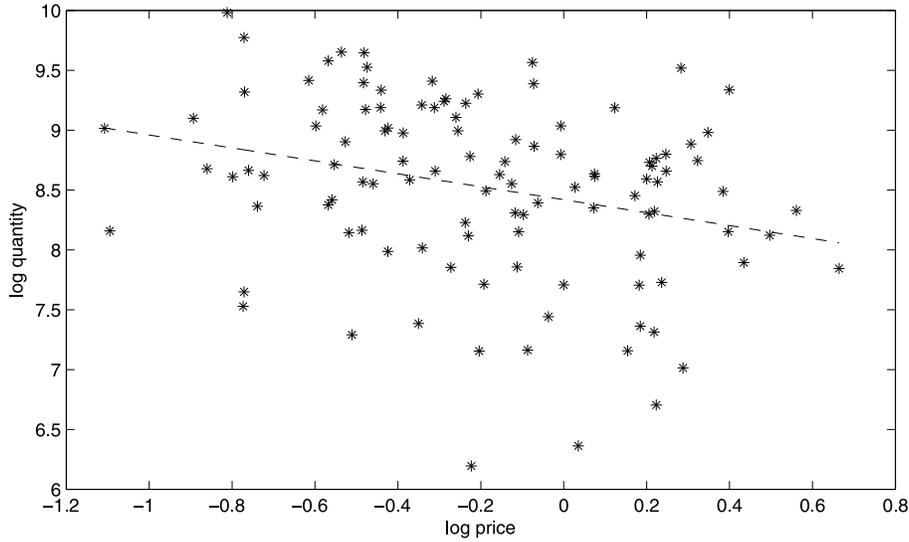}

\caption{Scatterplot of log prices and log quantities.}\label{f1}
\end{figure*}

%s3.1 #&#
\subsection{Discussions in the Statistics Literature}

To show the level of frustration and confusion in the statistics
literature with these models, let me present some quotes.
In a comment on Pratt and Schlaifer (\citeyear{Pra}), \citet{Daw84} writes
``I despair of ever understanding the logic of simultaneous
equations well enough to tackle them,'' (page 24).
\citet{Cox92}
writes in a discussion on causality ``it seems
reasonable that models should be
specified in a way that would allow direct computer simulation
of the data$\ldots$\,. This, for example, precludes the use of $y_2$ as an
explanatory variable for $y_1$ if at the same time $y_1$ is an
explanatory variable for $y_2$'' (page 294).
This restriction appears to rule out the first model Haavelmo
considers, that is, equations (1.1) and (1.2) (\cite*{Haa43}, page 2):
\[
Y=aX+\epsilon_1, \quad X=bY+\epsilon_2
\]
%
%In fact, the comment by Cox
(see also Haavelmo, \citeyear{Haa44}). In fact, the comment by Cox
appears to rule out all simultaneous
equations models of the type studied by economists.
%I will show
%, which often appears to be exactly what simultaneous equations models
%do.
\citet{Hol}, in comment on structural equation methods in econometrics,
writes ``why should [this disturbance] be
independent of [the instrument]$\ldots$ when the very
definition of [this disturbance] involves [the
instrument],''
(page 460).
Freedman writes
``Additionally, some
variables are taken to be exogenous (independent of the disturbance terms)
and some endogenous (dependent on the disturbance terms). The rationale
is seldom clear, because---among other things---there is seldom any very
clear description of what the disturbance terms mean, or where they come
from'' (\cite*{Fre06}, page 699).
%Making the example specific sheds some light on this. It also
%addresses the question
% \citet{Gel09} raises about the {\it ad hoc} nature of the
%instruments. For example, \citet{Gel09} quotes a colleague as claiming
%that researchers first find a good instrument and then the question
%that the instrument answers: ``What Angrist and his colleaguesdo is to
%find the instrument first, and then they go from there. They might see
%somethingin the newspaper or hear something on the radio and think,
%Hey--there is a natural experiment--it could make a good instrument!
%And then they go from there.'' (\cite*{Gel09}, p. 318) In the example
%the question is clear, and the instruments follow naturally from there.

%A key feature of supply and demand models is the presence of two (sets
%of) agents. The two agents have different objectives, and outcomes are
%determined by the interactions of the two agents.

%s3.2 #&#
\subsection{The Market for Fish}

The specific example I will use in this section is the market for
whiting (a particular white fish, often used in fish sticks) traded at
the Fulton fish market in New York City. Whiting was sold at the Fulton
fish market at the time by a small number of dealers to a large number
of buyers. Kathryn Graddy collected data on quantities and prices of
whiting sold by a particular trader at the Fulton fish market on 111
days between December 2, 1991, and May 8, 1992 (Graddy, \citeyear{Gra95}, \citeyear{Gra96};
\cite*{AngGraImb00}). I will take as the unit of analysis
a day, and interchangeably refer to this as a market. Each day, or
market, during the period covered in this data set, indexed by
$t=1,\ldots,111$, a number of pounds of whiting are sold by this
particular trader, denoted by $Q_t^\obs$. Not every transaction on the
same day involves the same price, but to focus on the essentials I will
aggregate the total amount of whiting sold and the total amount of
money it was sold for, and calculate a price per pound (in cents) for
each of the 111 days, denoted by $P_t^\obs$.
Figure~\ref{f1} presents a scatterplot of the observed log price and log
quantity data.
The average quantity sold over the 111 days was 6335 pounds, with a
standard deviation of 4040 pounds, for an average of the average
within-day prices of 88 cts per pound and a standard deviation of 34 cts.
For example, on the first day of this period 8058 pounds were sold for
an average of 65 cents, and the next day 2224 pounds were sold for an
average of 100 cents.
Table~\ref{summ_stats_fish} presents averages of log prices and log
quantities for the fish data.

%t1
%t1 #&#
\begin{table*}[b]
\tablewidth=13.5cm
\caption{Fulton fish market data ($N=111$)}
\label{summ_stats_fish}
\begin{tabular*}{\tablewidth}{@{\extracolsep{\fill}}ld{3.0}d{2.2}ccc@{}}
\hline
 &     & \multicolumn{2}{c}{\textbf{Logarithm of price}}                      & \multicolumn {2}{c@{}}{\textbf{Logarithm of quantity}}
 \\[-4pt]
 &&\multicolumn{2}{l}{\rule{4.14cm}{1pt}} &
 \multicolumn{2}{l@{}}{\rule{4.14cm}{1pt}}  \\
%  \ccline{3-4,5-6}
 &\multicolumn{1}{c}{\multirow{2}{55pt}[10pt]{\textbf{Number of observations}}} &   \multicolumn{1}{c}{\textbf{Average}}                                & \textbf{Standard deviation} & \textbf{Average}
 & \textbf{Standard deviation} \\
\hline
All        & 111 & -0.19 & (0.38) & 8.52 & (0.74) \\[3pt]
Stormy     & 32  & 0.04  & (0.35) & 8.27 & (0.71) \\
Not-stormy & 79  & -0.29 & (0.35) & 8.63 & (0.73) \\[3pt]
Stormy     & 32  & 0.04  & (0.35) & 8.27 & (0.71) \\
Mixed      & 34  & -0.16 & (0.35) & 8.51 & (0.77) \\
Fair       & 45  & -0.39 & (0.37) & 8.71 & (0.69) \\
\hline
\end{tabular*}
\end{table*}

Now suppose we are interested in predicting the effect of a tax in this
market. To be specific, suppose the government is considering imposing
a $100\times r$\% tax (e.g., a 10\% tax) on all whiting sold,
but before doing so it wishes to predict the average percentage change
in the quantity sold as a result of the tax. We may formalize that by
looking at the average effect on the logarithm of the quantity, $\tau
=\mathbb{E}[\ln Q_t(r)-\ln Q_t(0)]$, where $Q_t(r)$ is the quantity
traded in market/day $t$ if the tax rate were set at $r$. The problem,
substantially worse than in the standard causal inference setting where
for some units we observe one of the two potential outcomes and for
other units we observe the other potential outcome, is that in all 111
markets we observe the quantity traded at tax rate $0$, $Q_t^\mathrm{obs}=Q_t(0)$, and we never see the quantity traded at the tax rate
contemplated by the government, $Q_t(r)$.
% By necessity we need to make statements, from data on $Q_t(0)$ only,
%about the causal estimand $\tau$. More specifically,
Because only $\mathbb{E}[\ln Q_t(0)]$ is directly estimable from data
on the quantities we observe, the question is how to draw inferences
about $\mathbb{E}[\ln Q_t(r)]$.

A naive approach would be to assume that a tax increase by 10\% would
simply raise prices by 10\%.
If one additionally is willing to make the unconfoundedness assumption
that prices can be viewed as set independently of market conditions on
a particular day, it follows that those markets after the introduction
of the tax where the price net of taxes is \$1.00 would on average be
like those markets prior to the introduction of the 10\% tax where the
price was \$1.10. Formally, this approach assumes that
%
%e3.1 #&#
\begin{eqnarray}
\label{model1} && \mathbb{E}\bigl[\ln Q_t(r)|P^\obs
_t=p\bigr]
\nonumber
\\[-8pt]
\\[-8pt]
&& \quad =\mathbb{E}\bigl[\ln Q_t(0)|P^\obs_t=(1+r)
\times p\bigr],
\nonumber
\end{eqnarray}
implying that
\begin{eqnarray*}
&& \mathbb{E}\bigl[\ln Q_t(r)-\ln Q_t(0)|P^\obs_t=p
\bigr]
\\
&& \quad =\mathbb{E}\bigl[\ln Q_t^\obs
|P^\obs_t=(1+r)\times p\bigr]
\\
&& \qquad {}-\mathbb{E}\bigl[\ln Q_t^\obs|P^\obs_t=
p\bigr]
\\
&& \quad \approx\mathbb{E}\bigl[\ln Q_t^\obs|\ln
P^\obs_t=r+\ln p\bigr]
\\
&& \qquad {}-\mathbb{E}\bigl[\ln Q_t^\obs|\ln
P^\obs_t= \ln p\bigr].
\end{eqnarray*}
The last quantity is often estimated using linear regression methods.
Typically, the regression function is assumed to be
linear in logarithms with constant coefficients,
%
%e3.2 #&#
\begin{equation}
\label{regression} \ln Q_t^\obs= \alpha^{\mathrm{ls}}+
\beta^{\mathrm{ls}}\times \ln P_t^\obs+
\varepsilon_t.
\end{equation}
Ordinary least squares estimation with the Fulton fish market data
collected by Graddy leads to
\[
\begin{array} {ccccc} \widehat{\ln Q_t^\obs}=&
8.42&-&0.54& \times \ \ln P_t^\obs.
\\
& (0.08)&& (0.18)\end{array}
\]
The estimated regression line is also plotted in Figure~\ref{f1}.
Interestingly, this is what \citet{Wor27} calls the ``statistical `demand curve',''
as opposed to the concept of a demand curve in economic theory. This
simple regression, in combination with the assumption embodied in (\ref{model1}), suggests that the quantity traded would go down, on average,
by 5.4\% in response to a 10\% tax.
\[
\hat\tau=-0.054 \quad (\mbox{s.e. }0.018).
\]
Why does this answer, or at least the method in which it was derived,
not make any sense to an economist? The answer assumes that prices can
be viewed as independent of the potential quantities traded, or, in
other words, unconfounded.
This assignment mechanism is unrealistic. In reality, it is likely the
markets/days, prior to the introduction of the tax, when the price was
\$1.10 were systematically different from those where the price was \$1.00. From an economists' perspective, the fact that the price was
\$1.10 rather than \$1.00 implies that market conditions \textit{must} have
been different, and it is likely that these differences are directly
related to the potential quantities traded. For example, on days where the price
was high there may have been more buyers, or buyers may have been
interested in buying larger quantities, or there may have been less
fish brought ashore.
%it is likely that both buyers and sellers will respond systematically
%to the incentives that the introduction of the tax creates.
In order to predict the effect of the tax, we need to think about the
responses of both buyers and sellers to changes in prices, and about
the determination of prices. This is where economic theory comes in.

%s3.3 #&#
\subsection{The Supply of and Demand for Fish}

So, how do economists go about analyzing questions such as this one if
not by regressing quantities on prices? The starting point for
economists is to think of an economic model for the determination of
prices (the treatment assignment mechanism in Rubin's potential outcome
terminology). The first part of the simplest model an economist would
consider for this type of setting is a pair of functions, the demand
and supply functions. Think of the buyers coming to the Fulton
fishmarket on a given market/day (say, day $t$) with a demand function
$Q^d_t(p)$. This function tells us, for that particular morning, how
much fish all buyers combined would be willing to buy if the price on
that day were $p$, for any value of $p$. This function is conceptually
exactly like the potential outcomes set up commonly used in causal
inference in the modern literature. It is more complicated than the
binary treatment case with two potential outcomes, because there is a
potential outcome for each value of the price, with more or less a
continuum of possible price values, but it is in line with continuous
treatment extensions such as those in \citet{GilRob01}. Common
sense, and economic theory, suggests that this demand function is a
downward sloping function: buyers would likely be willing to buy more
pounds of whiting if it were cheaper. Traditionally, the demand
function is specified parametrically, for example, linear in logarithms:
%
%e3.3 #&#
\begin{equation}
\label{demand} \ln Q^d_t(p)=\alpha^d+
\beta^d\times \ln p+\varepsilon^d_t,
\end{equation}
where $\beta^d$ is the price elasticity of demand.
This equation is \textit{not} a regression function like (\ref
{regression}). It is interpreted as a \textit{structural equation} or
behavioral equation, and in the treatment effect literature
terminology, it is a model for the potential outcomes.
Part of the confusion between the model for the potential outcomes in
(\ref{demand}) and the regression function in (\ref{regression}) may
stem from the traditional notation in the econometrics literature where
the same symbol (e.g.,\vspace*{1pt} $Q_t$) would be used for the observed
outcomes ($Q^\obs_t$ in our notation) and the potential outcome
function [$Q^d_t(p)$ in our notation], and the same symbol (e.g.,
$P_t$) would be used for the observed value of the treatment ($P^\obs_t$ in our notation) and the argument in the potential\vadjust{\goodbreak} outcome function
($p$~in our notation). Interestingly, the pioneers in this literature,
Tinbergen (\citeyear{Tin30}) and \citet{Haa43}, \textit{did} distinguish between
these concepts in their notation, but the subsequent literature on
simultaneous equations dropped that distinction and adopted a notation
that did not distinguish between observed and
%potential outcomes. My
%view
potential outcomes. For a historical perspective see
Christ (\citeyear{Chr94}) and Stock and Trebbi (\citeyear{StoTre03}). My view
is that dropping this distinction was merely incidental, and that
implicitly the interpretation of the simultaneous equations models
remained that in terms of potential outcomes.\footnote{As a reviewer
pointed out, once one views simultaneous equations in terms of
potential outcomes, there is a natural normalization of the equations.
This suggests that perhaps the discussions of issues concerning
normalizations of equations in simultaneous equations models (e.g.,
Basmann, \citeyear{Bas63N1}, \citeyear{Bas63N2}, \citeyear{Bas65}; \cite*{Hil90}) implicitly rely on a
different interpretation, for example, thinking of the endogeneity
arising from measurement error. Throughout this discussion, I will
interpret simultaneous equations in terms of potential outcomes,
viewing the realized outcome notation simply as obscuring that.}

Implicit (by the lack of a subscript on the coefficients) in the
specification of the demand function in (\ref{demand}) is the strong
assumption that the effect of a unit change in the logarithm of the price
(equal to $\beta^d$)
is the same for all values of the price, and that the effect is the
same in all markets. This is clearly a very strong assumption, and the
modern literature on simultaneous equations (see \cite*{Mat07} for an
overview) has developed less restrictive specifications allowing for
nonlinear and nonadditive effects while maintaining identification.
The unobserved component in the demand function, denoted by
$\varepsilon^d_t$, represents unobserved determinants of the demand on
any given day/market: a particular buyer may be sick on a particular
day and not go to the market, or may be expecting a client wanting to
purchase a large quantity of whiting.
We can normalize this unobserved component to have expectation zero,
where the expectation is taken over all markets or days:
\[
\mathbb{E}\bigl[\ln Q^d_t(p)\bigr]=
\alpha^d+\beta^d\times\ln p.
\]
The interpretation of this expectation is subtle, and again it is part
of the confusion that sometimes arises. Consider the expected demand at
$p=1$, $\mathbb{E}[\ln Q^d_t(1)]$, under the linear
specification in (\ref{demand}) equal to $\alpha^d+\beta^d\cdot\ln
(1)=\alpha^d$. This $\alpha^d$ is the average of all demand
functions, evaluated at price equal to \$1.00, irrespective of what the
actual price in the market is, where the expectation is taken over \textit{all} markets. It is \textit{not}, and this is key, the {conditional}
expectation of the observed quantity in markets where the observed
price is equal to \$1.00 (or which is the same the demand function at 1
in those markets), which is $\mathbb{E}[\ln Q^\obs_t|\mbox{$P^\obs
_t=1$}]=\mathbb{E}[\ln Q^d_t(1)|P^\obs_t=1]$. %, for example, under a
%linear specification as in (\ref{regression}), equal to $\alpha^\mbox{%ls}+\beta^\mbox{ls}\cdot\ln(1)=\alpha^{\rm ls}$.
The original Tinbergen and Haavelmo notation and the modern potential
outcome version is helpful in making this distinction, compared to the
sixties econometrics textbook notation.\footnote{Other notations have
been recently proposed to stress the difference between the conditional
expectation of the observed outcome and the expectation of the
potential outcome.
\citet{Pea00} writes the
%terminology $\mathbb{E}\left[\ln Q^d_t(1)]$ is the
expected demand when the price is \textit{set} to \$1.00 as $\mathbb
{E}[\ln Q^d_t|\operatorname{do}(P_t=1)]$, rather than conditional
on the price being observed to be \$1.00.
\citet{HerRob06} write this average potential outcome as
$\mathbb{E}[\ln Q^d_t(P_t=1)]$, whereas
\citet{LauRic02} write it as $\mathbb{E}[\ln Q^\obs
_t\parallel P^\obs_t=1]$ where the double $\parallel$ implies conditioning by
intervention. }

Similar to the demand function, the supply function $Q^s_t(p)$
represents the quantity of whiting the sellers collectively are willing
to sell at any given price $p$, on day $t$. Here, common sense would
suggest that this function is sloping upward: the higher the price, the
more the sellers are willing to sell.
As with the demand function, the supply function is typically specified
parametrically with constant coefficients:
%
%e3.4 #&#
\begin{equation}
\label{supply} \ln Q^s_t(p)=\alpha^s+
\beta^s\times \ln p+\varepsilon^s_t,
\end{equation}
where $\beta^s$ is the price elasticity of supply.
Again we can normalize the expectation of $\varepsilon^s_t$ to zero
(where the expectation is taken over markets), and write
\[
\mathbb{E}\bigl[\ln Q^s_t(p)\bigr]=
\alpha^s+\beta^s\times\ln p.
\]
Note that the $\varepsilon^d_t$ and $\varepsilon^s_t$ are not assumed
to be independent in general, although in some applications that may be
a reasonable assumption. In this specific example, $\varepsilon_t^d$
may represent random variation in the set or number of buyers coming to
the market on a particular day, and $\varepsilon^s_t$ may represent
random variation in suppliers showing up at the market and in their
ability to catch whiting during the preceding days. These components
may well be uncorrelated, but there may be common components, for
example, in traffic conditions around the market that make it difficult
for both suppliers and buyers to come to the market.

%s3.4 #&#
\subsection{Market Equilibrium}

Now comes the second part of the simple economic model, the
determination of the price, or, in the terminology of the treatment
effect literature, the assignment mechanism.
The conventional\vadjust{\goodbreak} assumption in this type of market is that the price
that is observed, that is, the price at which the fish is traded in
market/day $t$, is the (unique) market clearing price at which demand
and supply are equal. In other words, this is the price at which the
market is in \textit{equilibrium}, denoted by $P^\obs_t$. This
equilibrium price solves
%
%e3.5 #&#
\begin{equation}
\label{equilibrium} Q^d_t\bigl(P_t^\obs
\bigr)=Q^s_t\bigl(P_t^\obs\bigr).
\end{equation}
The observed quantity on that day, that is the quantity actually
traded, denoted by $Q^\obs_t$, is then equal to the demand function at
the equilibrium price (or, equivalently, because of the equilibrium
assumption, the supply function at that price):
%
%e3.6 #&#
\begin{equation}
\label{q_equi} Q^\obs_t=Q^d_t
\bigl(P_t^\obs\bigr)=Q^s_t
\bigl(P_t^\obs\bigr).
\end{equation}
Assuming that the demand function does slope downward and the supply
function does slope upward, and both are linear in logarithms, the
equilibrium price exists and is unique, and we can solve for the
observed price and quantities in terms of the parameters of the model
and the unobserved components:
\begin{eqnarray*}
\ln P^\obs_t &=& \frac{\alpha^d-\alpha^s}{\beta^s-\beta^d}+ \frac{\varepsilon^d_t-\varepsilon^s_t}{\beta^s-\beta^d}
\quad \mbox{and}
\\
\ln Q^\obs_t &=& \frac{\beta^s\cdot\alpha^d-\beta^d\cdot\alpha
^s}{\beta^s-\beta^d}+ \frac{\beta^s\cdot\varepsilon^d_t-\beta^d\cdot\varepsilon
^s_t}{\beta^s-\beta^d}.
\end{eqnarray*}
For economists, this is a more plausible model for the determination of
realized prices and quantities than the model that assumes prices are
independent of market conditions. It is not without its problems
though. Chief among these from our perspective is the complication
that, just as in the Roy model, we cannot necessarily infer the values
of the unknown parameters in this model even if we have data on
equilibrium prices and quantities $P^\obs_t$ and $Q^\obs_t$ for many markets.

Another issue is how buyers and sellers arrive at the equilibrium price.
There is a theoretical economic literature addressing this question.
Often the idea is that there is a sequential process of buyers making
bids, and suppliers responding with offers of quantities at those
prices, with this process repeating itself until it arrives at a price
at which supply and demand are equal. In practice, economists often
refrain from specifying the details of this process and simply assume
that the market is in equilibrium. If the process is fast enough, it
may be reasonable to ignore the fact the specifics of the process and
analyze the data as if equilibrium was instantaneous.\footnote{See
\citet{ShaShu77} and \citet{Gir03}, and for some experimental
evidence, Plott and Smith (\citeyear{PloSmi87}) and \citet{Smi}.}
A related issue is whether this model with an equilibrium prices that
equates supply and demand is a reasonable approximation to the actual
process that determines prices and quantities. In fact, Graddy's data
contains information showing that the seller would trade at different
prices on the same day, so strictly speaking this model does not hold.
There is a long tradition in economics, however, of using such models
as approximations to price determination and we will do so here.

Finally, let me connect this to the textbook discussion of supply and
demand models. In many textbooks, the demand and supply equations would
be written directly in terms of the observed (equilibrium) quantities
and prices as
%
%e3.7 #&#
%e3.8 #&#
\begin{eqnarray}
\label{wold1} Q^\obs_t &=& \alpha^s+
\beta^s\times\ln P^\obs_t+\varepsilon^s_t,
\\
\label{wold2} Q^\obs_t &=& \alpha^d+
\beta^d\times\ln P^\obs_t+\varepsilon^d_t.
\end{eqnarray}
This representation leaves out much of the structure that gives the
demand and supply function their meaning, that is, the demand equation
(\ref{demand}), the supply equation (\ref{supply}) and the
equilibrium condition (\ref{equilibrium}). As \citet{StrWol60}
write, ``Those who write such systems [(\ref{wold2}) and (\ref{wold2})] do not, however, really mean what they write, but introduce
an ellipsis which is familiar to economists'' (page 425), with the
ellipsis referring to the market equilibrium condition
%that is left out.
that is left out. See also Strotz (\citeyear{Str60}), Strotz and Wold (\citeyear{StrWol65}),
and Wold (\citeyear{Wol60})

%s3.5 #&#
\subsection{The Statistical Demand Curve}

Given this set up, let me discuss two issues. First,
let us explore, under this model, the interpretation of what \citet{Wor27} called the ``statistical demand curve.'' The covariance between
observed (equilibrium) log quantities and log prices is
$ \operatorname{cov}  (\ln Q_t^\obs,\ln P^{\obs}_t )=(\beta
^s\cdot\sigma^2_{d}+
\beta^d\cdot\sigma^2_{s}-\rho\cdot\sigma_d\cdot\sigma_s\cdot
(\beta^d+\beta^s))/( (\beta^s-\beta^d )^2)$,
where $\sigma_d$ and $\sigma_s$ are the standard deviations of
$\varepsilon_t^d$ and $\varepsilon_t^s$, respectively, and $\rho$ is
their correlation.
Because the variance of $\ln P^{\obs}_t$ is $(\sigma_s^2+\sigma
^2_d-2\cdot\rho\cdot\sigma_d\cdot\sigma_s)/(\beta^s-\beta
^d)^2$, it follows that
the regression coefficient in the regression of log quantities on log
prices is
\begin{eqnarray*}
&& \frac{\operatorname{cov}  (\ln Q_t^{\obs},\ln P^{\obs}_t )}{\operatorname{var} (\ln P^{\obs}_t )}
\\
&& \quad =\frac{\beta^s\cdot\sigma^2_{d}+
\beta^d\cdot\sigma^2_{s}-\rho\cdot\sigma_d\cdot\sigma_s\cdot
(\beta^d+\beta^s)}{\sigma_s^2+\sigma^2_d-2\cdot\rho\cdot\sigma
_d\cdot\sigma_s}.
\end{eqnarray*}
Working focuses on the interpretation of this relation between
equilibrium quantities and prices. Suppose that
the correlation between $\varepsilon^d_t$ and $\varepsilon^s_t$,
denoted by $\rho$, is zero. Then the regression coefficient is a
weighted average of the two slope coefficients of the supply and demand
function, weighted by the variances of the residuals:
\[
%&&
\frac{\operatorname{cov} (\ln Q_t^{\obs},\ln P^{\obs}_t )}{\operatorname{var} (\ln P^{\obs}_t )} %\\
%&& \quad
=\beta^s\cdot
\frac{\sigma^2_{d}}{\sigma_s^2+\sigma^2_d} +\beta^d\cdot\frac{
\sigma^2_{s}}{\sigma_s^2+\sigma^2_d}. %\end{eqnarray*}
\]
If $\sigma_d^2$ is small relative to $\sigma^2_s$, then we estimate
something close to the slope of the demand function, and if $\sigma
_s^2$ is small relative to $\sigma^2_d$, then we estimate something
close to the slope of the supply function. In general, however, as
Working stresses, the ``statistical demand curve'' is not informative
about the demand function (or about the supply function); see also
\citet{LeaN1}.

%s3.6 #&#
\subsection{The Effect of a Tax Increase}

The
second question is how this model with supply and demand functions and
a market clearing price helps us answer the substantive question of
interest. The specific question considered is
the effect of the tax increase on the average quantity traded. In a
given market, let $p$ be the price sellers receive per pound of
whiting, and let $\tilde{p}=p\times(1+r)$ the price buyers pay after
the tax has been imposed. The key assumption is that the only way
buyers and sellers respond to the tax is through the effect of the tax
on prices: they do not change how much they would be willing to buy or
sell at any given price, and the process that determines the
equilibrium price does not change. The technical econometric term for
this is that the demand and supply functions are \textit{structural} or
\textit{invariant} in the sense that they are not affected by changes in
the treatment, taxes in this case. This may not be a perfect
assumption, but certainly in many cases it is reasonable: if I have to
pay \$1.10 per pound of whiting, I probably do not care whether 10 cts
of that goes to the government and \$1 to the seller, or all of it goes
to the seller. If we are willing to make that assumption, we can solve
for the new equilibrium price and quantity. Let $P_t(r)$ be the new
equilibrium price [net of taxes, that is, the price sellers receive,
with $(1+r)\cdot P_t(r)$ the price buyers pay], given a tax rate $r$,
with in our example $r=0.1$. This price solves
\[
Q^d_t\bigl(P_t(r)\times(1+r)
\bigr)=Q^s_t\bigl(P_t(r)\bigr).
\]
Given the log linear specification for the demand and supply functions,
this leads to
\[
\ln P_t(r)=\frac{\alpha^d-\alpha^s}{\beta^s- \beta^d}+ \frac{\beta^d\times\ln(1+r)}{\beta^s-\beta^d}+
\frac{\varepsilon^d_t-\varepsilon^s_t}{\beta^s- \beta^d}.
\]
The result of the tax is that the average of the logarithm of the price
that sellers receive with a positive tax rate $r$ is less than what
they would have received in the absence of the tax rate:
\begin{eqnarray*}
\mathbb{E}\bigl[\ln P_t(r)\bigr] & =& \frac{\alpha^d-\alpha^s}{\beta
^s-\beta^d}+
\frac{\beta^d\times\ln(1+r)}{\beta^s-\beta^d}
\\
& \leq & \frac{\alpha^d-\alpha^s}{\beta^s- \beta^d}= \mathbb{E}\bigl[\ln P_t(0)\bigr].
\end{eqnarray*}
(Note that $\beta^d<0$.)
On the other hand,
the buyers will pay more on average:
\begin{eqnarray*}
\mathbb{E}\bigl[\ln\bigl((1+r)\cdot P_t(r)\bigr)\bigr] &=&
\frac{\alpha
^d-\alpha^s}{\beta^s- \beta^d}+\frac{\beta^s\times\ln
(1+r)}{\beta^s-\beta^d}
\\
& \geq & %(1+r)\times\frac{\alpha^d-\alpha^s}{\beta^s-(1+r)\times\beta^d}
%=(1+r)\times
\mathbb{E}\bigl[\ln P_t(0)\bigr].
\end{eqnarray*}
The quantity traded after the tax increase is
\begin{eqnarray*}
\ln Q_t(r) &= & \frac{\beta^s\cdot\alpha^d-\beta^d\cdot\alpha
^s}{\beta^s-\beta^d} +\frac{\beta^s\cdot\beta^d\cdot\ln(1+r)}{\beta^s-\beta^d}
\\
&&{}+ \frac{\beta^s\cdot\varepsilon^d_t-\beta^d\cdot\varepsilon
^s_t}{\beta^s-\beta^d},
\end{eqnarray*}
which is less than the quantity that would be traded in the absence of
the tax increase. The
causal effect is
\[
\ln Q_t(r)-\ln Q_t(0)= \frac{\beta^s\cdot\beta^d\cdot\ln(1+r)}{\beta^s-\beta^d},
\]
the same in all markets, and
proportional to the supply and demand elasticities and, for small $r$,
proportional to the tax.
%with the average causal effect of the $100\times r$\% tax increase
%equal to
%.\]
What should we take away from this discussion? There are three points.
First, the regression coefficient in the regression of log quantity on
log prices does not tell us much about the effect of new tax. The sign
of this regression coefficient is ambiguous, depending on the variances
and covariance of the unobserved determinants of supply and demand.
Second, in order to predict the magnitude of the effect of a new tax we
need to learn about the demand and supply functions separately, or in
the econometrics terminology, \textit{identify} the supply and demand
functions. Third, observations on equilibrium prices and quantities by
themselves do not identify these functions.

%s3.7 #&#
\subsection{Identification with Instrumental Variables}

Given this identification problem, how \textit{do} we identify the demand
and supply functions? This is where instrumental variables enter the discussion.
To identify the demand function, we look for determinants of the
supply of whiting that do not affect the demand for whiting, and,
similarly, to identify the supply function we look for determinants of
the demand for whiting that do not affect the supply. In this specific
case, Graddy (\citeyear{Gra95}, \citeyear{Gra96}) assumes that weather conditions at sea on the
days prior to market $t$, denoted by $Z_t$, affect supply but do not
affect demand. Certainly, it appears reasonable to think that weather
is a direct determinant of supply: having high waves and strong winds
makes it harder to catch fish.
On the other hand, there does not seem to be any reason why demand on
day $t$, at a given price $p$, would be correlated with wave height or
wind speed on previous days. This assumption may be made more plausible
by conditioning on covariates. For example, if one is concerned that
weather conditions on land affect demand, one may wish to condition on
those, and only look at variation in weather conditions at sea given
similar weather conditions on land as an instrument.
Formally, the key assumptions are that
\[
Q_t^d(p) \perp Z_t \quad \mbox{and}\quad
Q_t^s(p) \not\perp Z_t,
\]
possibly conditional on covariates.
If both of these conditions hold,
we can use weather conditions as an instrument.

How do we exploit these assumptions? The traditional approach is to
generalize the functional form of the supply function to explicitly
incorporate the effect of the instrument on the supply of whiting. In
our notation,
\[
\ln Q^s_t(p,z)=\alpha^s+\beta^s
\times\ln p+\gamma^s\times z+\varepsilon^s_t.
\]
The demand function remains unchanged, capturing the fact that demand
is not affected by the instrument:
\[
\ln Q^d_t(p,z)=\alpha^d+\beta^d
\times\ln p+\varepsilon^d_t.
\]
We assume
that the unobserved components of supply and demand are independent of
(or at least uncorrelated with) the weather conditions:
\[
\bigl(\varepsilon_t^d,\varepsilon^s_t
\bigr) \perp Z_t.
\]
The equilibrium price $P^\obs_t$ is the solution for $p$ in the equation
\[
Q^d(p,Z_t)=Q^s_t(p,Z_t),
\]
which, in combination with the log linear specification for the demand
and supply functions, leads to
\[
\ln P^{\obs}_t=\frac{\alpha^d-\alpha^s}{\beta^s-\beta^d}+ \frac{\varepsilon^d_t-\varepsilon^s_t}{\beta^s-\beta^d}-
\frac{\gamma^s\cdot Z_t}{\beta^s-\beta^d}
\]
and
\begin{eqnarray*}
\ln Q^{\obs}_t &=& \frac{\beta^s\cdot\alpha^d-\beta^d\cdot\alpha
^s}{\beta^s-\beta^d}+ \frac{\beta^s\cdot\varepsilon^d_t-\beta^d\cdot\varepsilon
^s_t}{\beta^s-\beta^d}
\\
&& {}- \frac{\gamma^s\cdot\beta^d\cdot Z_t}{\beta^s-\beta^d}.
\end{eqnarray*}

Now consider the expected value of the equilibrium price and quantity
given the weather conditions:
%
%e3.9 #&#
\begin{eqnarray}
\label{eq:reduced1}
&& \mme\bigl[ \ln Q^{\obs
}_t|Z_t=z
\bigr]
\nonumber
\\[-8pt]
\\[-8pt]
&& \quad =\frac{\beta^s\cdot\alpha^d-\beta^d\cdot
\alpha^s}{\beta^s-\beta^d} - \frac{\gamma^s\cdot\beta^d}{\beta^s-\beta^d}\cdot z
\nonumber
\end{eqnarray}
and
%
%e3.10 #&#
\begin{equation}
\label{eq:reduced2} \qquad\quad\mme\bigl[ \ln P^{\obs}_t
|Z_t=z\bigr]=\frac{\alpha
^d-\alpha^s}{\beta^s-\beta^d}- \frac{\gamma^s}{\beta^s-\beta^d}\cdot z.
\end{equation}
Equations (\ref{eq:reduced1}) and
(\ref{eq:reduced2}) are what is called in econometrics the \textit{reduced form} of the simultaneous equations model.
It expresses the
\textit{endogenous} variables (those variables whose values are determined inside
the model, price and quantity in this example) in terms of the \textit{exogenous} variables (those variables whose values are not determined within the
model, weather conditions in this example).
The slope coefficients on the instrument in these reduced form
equations are what in randomized experiments with noncompliance would
be called the \textit{intention-to-treat} effects.
One can estimate the coefficients in the reduced form by least squares
methods. The key insight is that the ratio of the coefficients on the
weather conditions in the two regression functions,
$\gamma^s\cdot\beta^d/(\beta^s-\beta^d)$ in the quantity
regression and $\gamma^s/(\beta^s-\beta^d)$ in the price regression,
is equal to the slope coefficient in the demand function.

%f2
%f2 #&#
\begin{figure*}

\includegraphics{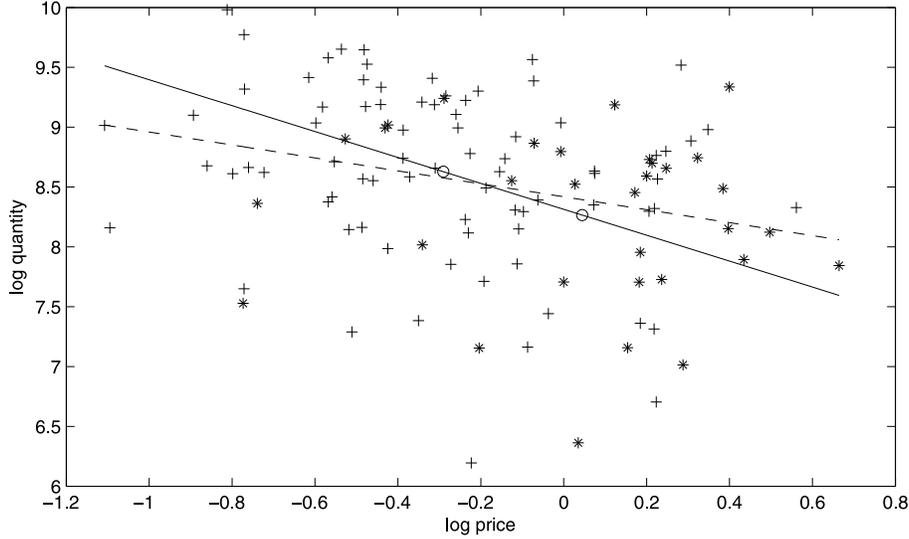}

\caption{Scatterplot of log prices and log quantities by weather conditions.}\label{fig2}
\end{figure*}

For some purposes, the reduced-form or intention-to-treat effects may
be of substantive interest. In the Fulton fish market example, people
attempting to predict prices and quantities under the current regime
may find these estimates of interest. They are of less interest to
policy makers contemplating the introduction of a new tax.
In simultaneous equations settings, the demand and supply functions are
viewed as \textit{structural} in the sense that they are not affected by
interventions in the market such as new taxes. As such they, and not
the reduced-form regression functions, are the key components of
predictions of market outcomes under new regimes.
This is somewhat different in many of the recent applications of
instrumental variables methods in the statistics literature in the
context of randomized experiments with noncompliance where the
intention-to-treat effects are traditionally of primary interest.

Let me illustrate this with the Fulton Fish Market data collected by
Graddy. For ease of illustration, let me simplify the instrument to a
binary one: the weather conditions are good for catching fish ($Z_t=0$,
fair weather, corresponding to low wind speed and low wave height) or
stormy ($Z_t=1$, corresponding to relatively strong winds and high
waves).\footnote{The formal definition I use, following \citet{AngGraImb00} is that stormy is defined as wind speed
greater than 18 knots in combination with wave height more than 4.5 ft,
and fair weather is anything else.} The price is the average daily
price in cents for one dealer, and the quantity is the daily quantity
in pounds. The two estimated reduced forms are
\[
\begin{array} {ccccc}\widehat{\ln Q}^\obs_t=&
8.63&-&0.36&\times Z_t
\\
& (0.08)&& (0.15) \end{array} %
\]
and
\[
\begin{array}{ccccc}
\widehat{\ln P}^\obs_t=& \!\!\!
-0.29&+&0.34&\times Z_t.
\\
& \!\!\!\hphantom{-}(0.04)&& (0.07) \end{array} %
\]
Hence, the instrumental variables estimate of the slope of the demand
function is
\[
\hat\beta^d=\frac{-0.36}{0.34}=-1.08\quad (\mbox{s.e. }0.46).
\]
Another, perhaps more intuitive way of looking at these estimates is to
consider the location of the average log quantity and average log price
separately by weather conditions. Figure~\ref{fig2} presents the scatter plot of
log quantity and log prices, with the stars indicating stormy days and
the plus signs indicating calm days.
On fair weather days the average log price is $-0.29$, and the average
log quantity is 8.6. On stormy days, the average log price is 0.04, and
the average log quantity is 8.3. These two loci are marked
by circles in Figure~\ref{fig2}.
On stormy days, the price is higher and the quantity traded is lower
than on fair weather days. This is used to estimate the slope of the
demand function.
The figure also includes the estimated demand function based on using
the indicator for stormy days as an instrument for the price: the
estimated demand function goes through the two points defined by the
average of the log price and log quantity for stormy and fair weather days.

With the data collected by Graddy, it is more difficult to point
identify the supply curve. The traditional route toward identifying the
supply curve would rely on finding an instrument that shifts demand
without directly affecting supply. Without such an instrument, we
cannot point identify the effect of the introduction of the tax on
quantity and prices. It is possible under weaker assumptions to find
bounds on these estimands (e.g., \cite*{LeaN1}; \cite*{Man03}), but
we do not pursue this here.

%s3.8 #&#
\subsection{Recent Research on Simultaneous Equations~Models}

The traditional econometric literature on simultaneous equations models
is surveyed in \citet{Hau83}. Compared to the discussion in the
preceding sections, this literature focuses on a more general case,
allowing for multiple endogenous variables and multiple instruments.
The modern econometric literature, starting in the 1980s, has relaxed
the linearity and additivity assumptions in specification (\ref
{demand}) substantially.
Key references to this literature are \citet{Bro83}, \citet{Roe88},
Newey and Powell (\citeyear{N03}), Chesher (\citeyear{Che03}, \citeyear{Che10}), \citet{BenBer06}, Matzkin (\citeyear{Mat03}, \citeyear{Mat07}),
\citet{AltMat05}, \citet{ImbNew09}, \citet{HodMam07},
\citet{Hor11} and \citet{HorLee07}.
\citet{Mat07} provides a recent survey of this technically demanding
literature.
This literature has continued to use the observed outcome notation,
making it more difficult to connect to the statistical literature.
Here, I briefly review some of this literature.
The starting point is a structural equation, in the potential outcome notation,
\[
Y_i(x)=\alpha+\beta\cdot x+\varepsilon_i
\]
and an instrument $Z_i$ that satisfies
\[
Z_i\perp\varepsilon_i\quad \mbox{and}\quad
Z_i\not\perp X_i.
\]
The traditional econometric literature would formulate this in the
observed outcome notation as
\[
Y_i=\alpha+\beta\cdot X_i+\varepsilon_i,
\quad Z_i\perp\varepsilon_i \quad \mbox{and}\quad
Z_i\not\perp X_i.
\]
There are a number of generalizations considered in the modern
literature. First, instead of assuming independence of the unobserved
component and the instrument, part of the current literature assumes
only that the conditional mean of the unobserved component given the
instrument is free of dependence on the instrument, allowing the
variance and other distributional aspects to depend on the value of the
instrument; see \citet{Hor11}. Another generalization of the linear
model allows for general nonlinear function forms of the type
\[
Y_i=g(X_i)+\varepsilon_i,\quad
Z_i\perp\varepsilon_i \quad \mbox{and} \quad
Z_i\not\perp X_i,
\]
where the focus is on nonparametric identification and estimation of
$g(x)$; see \citet{Bro83}, \citet{Roe88}, \citet{BenBer06}.
Allowing for even more generality, researchers have studied nonadditive
versions of these models with
\[
Y_i=g(X_i,\varepsilon_i),\quad
Z_i\perp\varepsilon_i \quad \mbox{and} \quad
Z_i\not\perp X_i,
\]
with $g(x,\varepsilon)$ strictly monotone in a scalar unobserved
component $\varepsilon$. In these settings, point identification often
requires strong assumptions on the support of the instrument and its
relation to the endogenous regressor and, therefore, researchers have
also explored bounds. See Matzkin (\citeyear{Mat03}, \citeyear{Mat07}, \citeyear{Mat08}) and \citet{ImbNew09}.

%s4 #&#
\section{A Modern Example: Randomized Experiments with Noncompliance
and Heterogenous Treatment Effects}
\label{section:modern}

In this section, I will discuss part of the modern literature on
instrumental variables methods that has evolved simultaneously in the
statistics and econometrics literature. I will do so in the context of
a second example.
On the one hand, concern arose in the econometric literature about the
restrictiveness of the functional form assumptions in the traditional
instrumental variables methods and in particular with the constant
treatment effect assumption that were commonly used in the so-called
selection models (\cite{Hec79}; \cite{HecRob}).
The initial results in this literature demonstrated the difficulties in
establishing point identification (\cite{Hec}; \cite{Man90}),
leading to the bounds approach developed by Manski (\citeyear{Man95}, \citeyear{Man03}). At the
same time, statisticians analyzed the complications arising from
noncompliance in randomized experiments (\cite{Rob}) and the merits
of encouragement designs (Zelen, \citeyear{Zel79}, \citeyear{Zel90}).
By adopting a common framework and notation in \citet{ImbAng94} and \citet{AngImbRub}, these literatures have
become closely connected and influenced each other substantially.

%s4.1 #&#
\subsection{\texorpdfstring{The McDonald, Hiu  and Tierney (\citeyear{McDHiuTie92}) Data}{The McDonald, Hiu  and Tierney (1992) Data}}

The canonical example in this literature is that of a randomized
experiment with noncompliance.
To illustrate the issues,
I will use here data previously analyzed in Hirano et al. (\citeyear{Hiretal00})
 and
McDonald, Hiu  and Tierney (\citeyear{McDHiuTie92}).
McDonald, Hiu  and Tierney (\citeyear{McDHiuTie92}) carried out a randomized experiment to
evaluate the effect of an influenza vaccination on flu-related hospital
visits. Instead of randomly assigning individuals to receive the
vaccination, the researchers randomly assigned physicians to receive
letters reminding them of the upcoming flu season and encouraging them
to vaccinate their patients. This is what Zelen (\citeyear{Zel79}, \citeyear{Zel90}) refers to
as an \textit{encouragement design}. I discuss this using the potential
outcome notation used for this particular set up in \citet{AngImbRub}, and in general sometimes referred to as the Rubin Causal
Model (\cite{Hol86}), although there are important antecedents in
Splawa-Neyman (\citeyear{Spl90}). I consider two distinct treatments: the first the
receipt of the letter, and second the receipt of the influenza
vaccination. Let $Z_i\in\{0,1\}$ be the indicator for the receipt of
the letter, and let $X_i\in\{0,1\}$ be the indicator for the receipt
of the vaccination. We start by postulating the existence of four
potential outcomes. Let $Y_i(z,x)$ be the potential outcome
corresponding to the receipt of letter equal to $Z_i=z$, and the
receipt of vaccination equal to $X_i=x$, for $z=0,1$ and $x=0,1$.
In addition, we postulate the existence of two potential outcomes
corresponding to the receipt of the vaccination as a function of the
receipt of the letter, $X_i(z)$, for $z=0,1$.
We observe for each unit in a population of size $N=2861$ the value of
the assignment, $Z_i$, the treatment actually received, $X_i^\obs
=X_i(Z_i)$ and the potential outcome corresponding to the assignment
and treatment received, $Y^\obs_i=Y_i(Z_i,X_i(Z_i))$.
Table~\ref{summ_stats_flu} presents the number of individuals for each
of the eight values of the triple $(Z_i,X^\obs_i,Y^\obs_i)$ in the
McDonald, Hiu  and Tierney data set.
It should be noted that the randomization in this experiment is at the
physician level. I do not have physician indicators and, therefore,
ignore the clustering. This will tend to lead to underestimation of the
standard errors.

%t2 #&#
\begin{table}
\caption{Influenza data ($N=2861$)}\label{summ_stats_flu}
\begin{tabular*}{\tablewidth}{@{\extracolsep{\fill}}lccd{4.0}@{}}
\hline
\textbf{Hospitalized for} & \multicolumn{1}{c}{\textbf{Influenza}} &  &
 \\
\textbf{flu-related reasons}         &         \textbf{vaccine}         &    \textbf{Letter}    &            \multicolumn{1}{c@{}}{\textbf{Number of}}           \\
$\boldsymbol{Y}_{\bolds{i}}^{\mathbf{obs}}$                  & $\bolds{X}^{\mathbf{obs}}_{\bolds{i}}$        & $\bolds{Z}_{\bolds{i}}$  & \multicolumn{1}{c@{}}{\textbf{individuals}}                      \\
 \hline
No                          & No                & No     & 1027                  \\
No                          & No                & Yes    & 935                   \\
No                          & Yes               & No     & 233                   \\
No                          & Yes               & Yes    & 422                   \\
Yes                         & No                & No     & 99                    \\
Yes                         & No                & Yes    & 84                    \\
Yes                         & Yes               & No     & 30                    \\
Yes                         & Yes               & Yes    & 31                    \\
\hline
\end{tabular*}
\end{table}

%s4.2 #&#
\subsection{Instrumental Variables Assumptions}
\label{assumptions}

There are four key of assumptions underlying instrumental variables
methods beyond the no-interference assumption or SUTVA, with different
versions for some of them.
I will introduce these assumptions in this section, and in Section~\ref{section:content} discuss their substantive content in the context of
some examples.
The first assumption concerns the assignment to the instrument $Z_i$,
in the flu example the receipt of the letter by the physician. The
assumption requires that the instrument is as good as randomly assigned:
%
%e4.1 #&#
%e4.2 #&#
\begin{eqnarray}
\label{eq:een}
&& Z_i \perp \bigl( Y_i(0,0),Y_i(0,1),Y_i(1,0),
\nonumber
\\
&&\hspace{48pt}  Y_i(1,1),X_i(0),X_i(1) \bigr)
\\
\eqntext{\mbox{(random assignment).}}
\end{eqnarray}
This assumption is often satisfied by design: if the assignment is
physically randomized, as the letter in the flu example and as in many
of the applications in the statistics literature (e.g., see the
discussion in \cite*{Rob}), it is automatically satisfied. In other
applications with observational data, common in the econometrics
literature, this assumption is more controversial. It can in those
cases be relaxed by requiring it to hold only within subpopulations
defined by covariates $V_i$, assuming the assignment of the instrument
is unconfounded:
%
%e4.3 #&#
\begin{eqnarray}
&& Z_i \perp \bigl( Y_i(0,0),Y_i(0,1),Y_i(1,0),
\nonumber\\
&&\hspace{48pt} Y_i(1,1),X_i(0),X_i(1) \bigr) |
V_i
\nonumber\\
\eqntext{(\mbox{unconfounded assignment given }V_i).}
\end{eqnarray}
This is identical to the generalization from random assignment to
unconfounded assignment in observational studies.
Either version of this assumption justifies the causal interpretation
of \textit{Intention-To-Treat} (ITT) effects, the comparison of outcomes
by assignment to the treatment. In many cases, these ITT effects are
only of limited interest, however, and this motivates the consideration
of additional assumptions that do allow the researcher to make
statements about the causal effects of the treatment of interest.
It should be stressed, however, that in order to draw inferences beyond
ITT effects, additional assumptions will be used; whether the resulting
inferences are credible will depend on the credibility of these assumptions.

The second class of assumptions limits or rules out completely direct
effects of the assignment (the receipt of the letter in the flu
example) on the outcome, other than through the effect of the
assignment on the receipt of the treatment of interest (the receipt of
the vaccine). This is the most critical, and typically most
controversial assumption underlying instrumental variables methods,
sometimes viewed as the defining characteristic of instruments. One way
of formulating this assumption is as
%e4.4 #&#
\begin{eqnarray}
&& Y_i(0,x)=Y_i(1,x)\quad \mbox{for } x=0,1, \mbox{for
all } i
\nonumber\\
\eqntext{\mbox{(exclusion restriction).}}
\end{eqnarray}
\citet{Rob} formulates a similar assumption as requiring that the
instrument is ``not an independent causal risk factor'' (\cite*{Rob},
page 119).
Under this assumption, we can drop the $z$ argument of the potential
outcomes and write the potential outcomes without ambiguity as $Y_i(x)$.
This assumption is typically a substantive one. In the flu example, one
might be concerned that the physician, in response to the receipt of
the letter, takes actions that affect the likelihood of the patient
getting infected with the flu other than simply administering the flu
vaccine. In randomized experiments with noncompliance, the exclusion
restriction is sometimes made implicitly by indexing the potential
outcomes only by the treatment $x$ and not the instrument $z$ (e.g., \cite*{Zel90}).

There are other, weaker versions of this assumption. Hirano et al. (\citeyear{Hiretal00}) use a stochastic version of the exclusion
restriction that only requires that the distribution of $Y_i(0,x)$ is
the same as the distribution of $Y_i(1,x)$. \citet{Man90} uses a weaker
restriction that he calls a \textit{level set restriction}, which requires
that the average value of $Y_i(0,x)$ is equal to the average value of
$Y_i(1,x)$. In another approach, Manski and Pepper (\citeyear{ManPep00}) consider
monotonicity assumptions that restrict the sign of $Y_i(1,x)-Y_i(0,x)$
across individuals without requiring that the effects are completely absent.

\citet{ImbAng94} combine the random assignment assumption and
the exclusion restriction by postulating the existence of a pair of
potential outcomes $Y_i(x)$, for $x=0,1$, and directly assuming that
\[
Z_i \perp \bigl(Y_i(0),Y_i(1) \bigr).
\]
A disadvantage of this formulation is that it becomes less clear
exactly what role randomization of the instrument plays.
Another version of this combination of the exclusion restriction and
random assignment assumption
does not require full independence, but assumes that the conditional
mean of $Y_i(0)$ and $Y_i(1)$ given the instrument is free of
dependence on the instrument. A~concern with such assumptions is that
they are functional form dependent: if they hold in levels, they do not
hold in logarithms unless full independence holds.

A third assumption that is often used, labeled \textit{monotonicity} by
\citet{ImbAng94}, requires that
\[
X_i(1)\geq X_i(0)\quad \mbox{for all } i \quad
\mbox{(monotonicity)},
\]
for all units. This assumption rules out the presence of units who
always do the opposite of their assignment [units with $X_i(0)=1$ and
$X_i(1)=0$], and is therefore also referred to as the \textit{no-defiance}
assumption (\cite{autokey15}).
It is implicit in the latent index models often used in econometric
evaluation models (e.g., Heckman and Robb, \citeyear{HecRob}).
In the randomized experiments such as the flu example, this assumption
is often plausible. There it requires that in response to the receipt
of the letter by their physician, no patient
is less likely to get the vaccine. \citet{Rob} makes this assumption
in the context of a randomized trial for the effect of AZT on AIDS, and
describes the assumption as ``often, but not always, reasonable''
(\cite*{Rob}, page~122).

Finally, we need the instrument to be correlated with the treatment, or
the instrument to be \textit{relevant} in the terminology of \citet{Phi89} and \citet{StaSto97}:
\[
X_i\not\perp Z_i.
\]
In practice, we need the correlation to be substantial in order to draw
precise inferences. A recent literature on \textit{weak instruments} is
concerned with credible inference in settings where this correlation
between the instrument and the treatment is weak; see \citet{StaSto97} and \citet{AndSto}.

The random assignment assumption and the exclusion restriction are
conveniently captured by the graphical model below, although the
monotonicity assumption does not fit in as easily. %, \textit{e.g.,}
%Figure 1 in \citet{HerRob06}.
The unobserved component $U$ has a direct effect on both the treatment
$X$ and the outcome $Y$ (captured by arrows from $U$ to $X$ and to~$Y$). The instrument $Z$ is not related to the unobserved component $U$
(captured by the absence of a link between $U$ and $Z$), and is only
related to the outcome $Y$ through the treatment $X$ (as captured by
the arrow from $Z$ to $X$ and an arrow from $X$ to $Y$, and the absence
of an arrow between $Z$ and $Y$).

%f3 #&#
\begin{figure}[h!]

\includegraphics{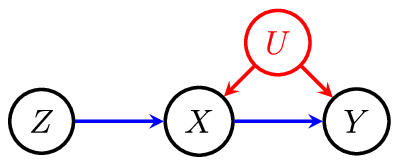}

\end{figure}

I will primarily focus on the case with all four assumptions
maintained, random assignment, the exclusion restriction, monotonicity
and instrument relevance, without additional covariates, because this
case has been the focus of, or a special case of the focus of, many
studies, allowing me to compare different approaches. Methodological
studies considering essentially this set of assumptions, sometimes
without explicitly stating instrument relevance, and sometimes adding
additional assumptions, include \citet{Rob}, \citet{Hec}, \citet{Man90},
\citet{ImbAng94}, \citet{AngImbRub}, \citet{RobGre96}, Balke and Pearl (\citeyear{autokey15}, \citeyear{BalPea97}), \citet{Gre}, Hern\'
an and Robins (\citeyear{HerRob06}), \citet{Rob94}, \citet{RobRot04},
\citet{VanGoe03}, Vansteelandt et al. (\citeyear{Vanetal11}), Hirano et al. (\citeyear{Hiretal00}),
%Tan (\citeyear{Tan06}, \citeyear{Tan10}) and others.
Tan (\citeyear{Tan06}, \citeyear{Tan10}), Abadie (\citeyear{Aba02}, \citeyear{Aba03}), Duflo,
Glennester and Kremer (\citeyear{DufGleKre07}), Brookhart et al. (\citeyear{Broetal06}),
Martens  et al. (\citeyear{Maretal06}), Morgan and
Winship (\citeyear{MorWin07}), and others.
Many more studies make the same assumptions in combination with a
constant treatment effect assumption.

The modern literature analyzed this setting from a number of different
approaches. Initially, the literature focused on the inability, under
these four assumptions, to identify the average effect of the
treatment. Some researchers, including prominently \citet{Man90}, \citet{autokey15} and \citet{Rob}, showed that although one could not
point-identify the average effect under these assumptions, there was
information about the average effect in the data under these
assumptions and they derived bounds for it. Another strand of the
literature, starting with \citet{ImbAng94} and \citet{AngImbRub} abandoned the effort to do inference for the overall
average effect, and focused on subpopulations for which the average
effect could be identified, the so-called compliers, leading to the
local average treatment effect. We discuss the bounds approach in the
next section (Section~\ref{bounds}) and the local average treatment
effect approach in Sections~\ref{types}--\ref{section:dowecare}.

%s4.3 #&#
\subsection{Point Identification {versus} Bounds}
\label{bounds}

In a number of studies, the primary estimand is the average effect of
the treatment, or the average effect for the treated:
%
%e4.5 #&#
\begin{eqnarray}
\label{ate} \tau & = &\mathbb{E}\bigl[Y_i(1)-Y_i(0)
\bigr]\quad \mbox{and}
\nonumber
\\[-8pt]
\\[-8pt]
\tau_t &=& \mathbb{E}\bigl[Y_i(1)-Y_i(0)|X_i=1
\bigr].
\nonumber
\end{eqnarray}
With only the four assumptions, random assignment, the exclusion
restriction, monotonicity, and instrument relevance \citet{Rob},
\citet{Man90} and \citet{autokey15} established that the
average treatment effect can often not be consistently estimated even
in large samples. In other words, that it is often \textit{not point-identified}.

Following this result, a number of different approaches have been taken.
\citet{Hec} showed that if the instrument takes on values such that
the probability of treatment given the instrument can be arbitrarily
close to zero and one, then the average effect is identified. This is
sometimes referred to as \textit{identification at infinity}.
\citet{Rob} also formulates assumptions that allow for point
identification, focusing on the average effect for the treated, $\tau
_t$. These assumptions restrict the average value of the potential
outcomes when not observed in terms of average outcomes that are
observed. For example, Robins formulates the condition that
\begin{eqnarray*}
&& \mathbb{E}\bigl[Y_i(1)-Y_i(0)|Z_i=1,X_i=1
\bigr]
\\
&& \quad =\mathbb{E}\bigl[Y_i(1)-Y_i(0)|Z_i=0,X_i=1
\bigr],
\end{eqnarray*}
which,
in combination with the random assignment and the exclusion
restriction, this allows for point identification of the average effect
for the treated. Robins also formulates two other assumptions,
including one where the effects are proportional to survival rates
$\mathbb{E}[Y_i(1)|Z_i=1,X_i=1]$ and $\mathbb{E}[Y_i(1)|Z_i=0,X_i=1]$
respectively, that also point-identifies the average effect for the
treated. However, Robins questions the applicability of these results
by commenting that ``it would be hard to imagine that there is
sufficient understanding of the biological mechanism$\ldots$ to have strong
beliefs that any of the three conditions$\ldots$ is more likely to hold
than either of the other two'' (\cite*{Rob}, page~122).

As an alternative to adding assumptions, \citet{Rob}, \citet{Man90}
and \citet{autokey15}, focused on the question what can be learned
about $\tau$ or $\tau_t$ given these four assumptions that do not
allow for point identification. Here, I focus on the case where the
three assumptions, random assignment, the exclusion restriction and
monotonicity % and instrument relevance,
are maintained
(without necessarily instrument relevance holding), although \citet{Rob} and \citet{Man90} also consider other combinations of
assumptions. For ease of exposition, I focus on the bounds for the
average treatment effect $\tau$ under these assumptions, in the case
where $Y_i(0)$ and $Y_i(1)$ are binary. Then
\begin{eqnarray*}
&& \mathbb{E}\bigl[Y_i(1)-Y_i(0)\bigr]
\\
&& \quad \in \bigl[-\bigl(1-\mme[X_i|Z_i=1]\bigr)\cdot
\mathbb{E}[Y_i|Z_i=1,X_i=0]
\\
&&\qquad \hspace*{2pt} {}+ \mme[Y_i|Z_i=1]-
\mme[Y_i|Z_i=0]
\\
&&\qquad \hspace*{2pt} {}+\mme[X_i|Z_i=0]\cdot \bigl(
\mathbb{E}[Y_i|Z_i=0,X_i=1]-1\bigr),
\\
&& \qquad \hspace*{2pt}\bigl(1-\mme[X_i|Z_i=1]\bigr)
\\
&&\qquad \hspace*{2pt} {}\cdot\bigl(1- \mathbb{E}[Y_i|Z_i=1,X_i=0]
\bigr)
\\
&&\qquad \hspace*{2pt} {}+ \mme[Y_i|Z_i=1]-
\mme[Y_i|Z_i=0]
\\
&&\qquad \hspace*{32pt} {}+\mme[X_i|Z_i=0]\cdot
\mathbb{E}[Y_i|Z_i=0,X_i=1] \bigr],
\end{eqnarray*}
which are known at the \textit{natural bounds}.
In this simple setting, this is a straightforward calculation. Work by
Manski (\citeyear{Man95}, \citeyear{Man03}, \citeyear{Man05}, \citeyear{Man07}),
\citet{Rob} and \citet{HerRob06} extends the partial identification approach to substantially
more complex settings.

For the McDonald--Hiu--Tierney flu data, the estimated identified set for
the population average treatment effect is
\[
\mathbb{E}\bigl[Y_i(1)-Y_i(0)\bigr]\in [ -0.24, 0.64].
\]
%
%There is obviously also uncertainty associated with these bounds.
There is a growing literature developing methods for establishing
confidence intervals for parameters in settings with partial
identification taking sampling uncertainty into account; see Imbens and
\citet{Man04} and \citet{CheHonTam07}.

%s4.4 #&#
\subsection{Compliance Types}
\label{types}

\citet{ImbAng94} and \citet{AngImbRub} take a
different approach. Rather than focusing on the average effect for the
population that is not identified under the three assumptions given in
Section~\ref{assumptions}, they focus on different average causal effects.
A first key step in the Angrist--Imbens--Rubin set up is that we can
think of four different compliance types defined by the pair of values
of $(X_i(0),X_i(1))$, that is, defined by how individuals would respond
to different assignments in terms of receipt of the treatment:\footnote
{\citet{FraRub02} generalize this notion of subpopulations
whose membership is not completely observed into their \textit{principal
stratification} approach; see also Section~\ref{sec72}.}
\[
T_i= \cases{ n \ (\never) & $\ifff X_i(0)=X_i(1)=0$,
\cr
c \ (\comp) & $\ifff X_i(0)=0, X_i(1)=1$,
\cr
d \ (
\defier) & $\ifff X_i(0)=1,X_i(1)=0$,
\cr
a \ (\always) &
$\ifff X_i(0)=X_i(1)=1$.}
\]
Given the existence of deterministic potential outcomes this
partitioning of the population into four subpopulations is simply a
definition.\footnote{Outside of this framework, the existence of these
four subpopulations would be an assumption.} It clarifies immediately
that it will be difficult to identify the average effect of the primary
treatment (the receipt of the vaccine) for the entire population:
never-takers and always-takers can only be observed exposed to a single
level of the treatment of interest, and thus for these groups any point
estimates of the causal effect of the treatment must be based on extrapolation.

We cannot infer without additional assumptions the compliance type of
any unit: for each unit we observe $X_i(Z_i)$, but the data contain no
information about the value of $X_i(1-Z_i)$. For each unit, there are
therefore two compliance types consistent with the observed behavior.
We can also not identify the proportion of
individuals of each compliance type without additional restrictions.
The monotonicity assumption implies that there are no defiers. This, in
combination with random assignment, implies that we can identify the
population shares of the remaining three compliance types. The
proportion of always-takers and never-takers are
\begin{eqnarray*}
\pi_a &=& \pr(T_i=a)=\pr(X_i=1|Z_i=0)
\quad \mbox{and}
\\
%The proportion of never-takers is
\pi_n &=& \pr(T_i=n)=\pr(X_i=0|Z_i=1),
\end{eqnarray*}
respectively,
and the proportion of compliers is the remainder:
\[
\pi_c=\pr(T_i=c)=1-\pi_a-
\pi_n.
\]
For the McDonald--Hiu--Tierney data these shares are estimated to be
\[
\hat\pi_a=0.189,\quad \hat\pi_n=0.692,\quad \hat
\pi_c=0.119,
\]
although, as I discuss in
Section~\ref{section:exclusion}, these shares may not be consistent
with the exclusion restriction.

%s4.5 #&#
\subsection{Local Average Treatment Effects}
\label{section_late}

If, in addition to monotonicity, we also assume that the exclusion
restriction holds, \citet{ImbAng94} and \citet{AngImbRub} show that the
\textit{local average treatment effect} or \textit{complier average causal
effect} is identified:
%
%e4.6 #&#
\begin{eqnarray}
\label{eq:late} \tau_{\mathrm{late}} &=& \mathbb{E}\bigl[ Y_i(1)-Y_i(0)
|T_i= c \bigr]
\nonumber
\\[-8pt]
\\[-8pt]
&=& \frac{\mathbb{E}[ Y_i|Z_i=1]-\mathbb{E}[
Y_i|Z_i=0]}{\mathbb{E}[ X_i|Z_i=1]-\mathbb
{E}[ X_i|Z_i=0]}.
\nonumber
\end{eqnarray}
The components of the right-hand side of this expression can be
estimated consistently from a random sample $(Z_i,X_i,Y_i)_{i=1}^N$.
For the McDonald--Hiu--Tierney data, this leads to
\[
\hat{\tau}_{\mathrm{late}}= -0.125\quad (\mbox{s.e. }0.090).
\]

Note that just as in the supply and demand example, the causal estimand
is the ratio of the intention-to-treat effects of the letter on
hospitalization and of the letter on the receipt of the vaccine. These
intention-to-treat effects
are
\begin{eqnarray*}
\widehat{\mbox{ITT}}_Y & =& -0.015 \quad (\mbox{s.e. } 0.011),
\\
\widehat{\mbox{ITT}}_X &=& \hat\pi_c= 0.119 \quad (
\mbox{s.e. }0.016),
\end{eqnarray*}
with the latter equal to the estimated proportion of compliers in the
population.

Without the monotonicity assumption, but maintaining the random
assignment assumption and the exclusion restriction, the ratio of ITT
effects still has a clear interpretation. In that case, it
is equal to a linear combination average of the effect of the treatment
for compliers and defiers:
%
%e4.7 #&#
\begin{eqnarray}
\label{eq:late1} \quad&& \frac{\mathbb{E}[ Y_i|Z_i=1]-\mathbb{E}[
Y_i|Z_i=0]}{\mathbb{E}[ X_i|Z_i=1]-\mathbb
{E}[ X_i|Z_i=0]}
\nonumber
\\
&& \quad =\frac{\operatorname{pr}(T_i=c)}{\operatorname{pr}(T_i=c)-\operatorname{pr}(T_i=d)}
\nonumber
\\
&& \qquad {}\cdot \mathbb{E}\bigl[ Y_i(1)-Y_i(0)
|T_i=c\bigr]
\\
&& \qquad {}- \frac{\operatorname{pr}(T_i=d)}{\operatorname{pr}(T_i=c)-\operatorname{pr}(T_i=d)}
\nonumber
\\
&& \quad\qquad {}\cdot \mathbb{E}\bigl[ Y_i(1)-Y_i(0)
|T_i=d\bigr].
\nonumber
\end{eqnarray}
This estimand has a clear interpretation if the treatment effect is
constant across all units, but if there is heterogeneity in the
treatment effects it is a weighted average with some weights negative.
This representation shows that if the monotonicity assumption is
violated, but the proportion of defiers is small relative to that of
compliers, the interpretation of the instrumental variables estimand is
not severely impacted.

%s4.6 #&#
\subsection{Do We Care About the Local Average Treatment Effect?}
\label{section:dowecare}

The local average treatment effect is an unusual estimand. It is an
average effect of the treatment for a subpopulation that cannot be
identified in the sense that there are no units whom we know for sure
to belong to this subpopulation, although there are some units whom we
know do not belong to it.
A more conventional approach is to start an analysis by clearly
articulating the object of interest, say the average effect of a
treatment for a well-defined population. There may be challenges in
obtaining credible estimates of this object of interest, and along the
way one may make more or less credible assumptions, but typically the
focus remains squarely on the originally specified object of interest.

Here, the approach appears to be quite different. We started off by
defining unit-level treatment effects for all units. We did not
articulate explicitly what the target estimand was. In the
McDonald--Hiu--Tierney influenza-vaccine application a natural estimand
might be the population average effect of the vaccine. Then, apparently
more or less by accident, the definition of the compliance types led us
to focus on the average effects for compliers. In this example, the
compliers were defined by the response in terms of the receipt of the
vaccine to the receipt of the letter. It appears difficult to argue
that this is a substantially interesting group, and in fact no attempt
was made to do so.

This type of example has led distinguished researchers both in
economics and in statistics to question whether and why one should care
about the local average treatment effect. The economist Deaton writes
``I find it hard to make any sense of the LATE [local average treatment
effect]''
(\cite*{Dea10}, page 430). Pearl similarly wonders ``Realizing that the
population averaged treatment effect
(ATE) is not identifiable in experiments marred by noncompliance, they have
shifted attention to a specific response type (i.e., compliers) for
which the causal
effect was identifiable, and presented the latter [the local average
treatment effect] as an approximation for ATE. $\ldots$ However, most
authors in this category do not state
explicitly whether their focus on a specific stratum is motivated by
mathematical
convenience, mathematical necessity (to achieve identification) or a genuine
interest in the stratum under \mbox{analysis}'' (\cite*{Pea11}, page 3).
Freedman writes ``In
many circumstances, the instrumental-variables estimator turns out to be
estimating some data-dependent average of structural parameters, whose
meaning would have to be elucidated'' (\cite*{Fre06}, pages 700--701).
Let me attempt to clear up this
%confusion. An instrumental
confusion. See also Imbens (\citeyear{Imb10}). An instrumental
variables
analysis is an analysis in a second-best setting. It would have been
preferable if one had been able to carry out a well-designed randomized
experiment. However, such an experiment was not carried out, and we
have noncompliance. As a result, we cannot answer all the questions we
might have wanted to ask. Specifically, if the noncompliance is
substantial, we are limited in the questions we can answer credibly and
precisely. Ultimately, there is only one subpopulation we can credibly
(point-)identify the average effect of the treatment for, namely, the compliers.

It may be useful to draw an analogy. Suppose a researcher is interested
in evaluating a medical treatment and suppose a randomized experiment
had been carried out to estimate the average effect of this new
treatment. However, the population of the randomized experiment
included only men, and the researcher is interested in the average
effect for the entire population, including both men and women. What
should the researcher do? I would argue that the researcher should
report the results for the men, and acknowledge the limitation of the
results for the original question of interest. Similarly, in the
instrumental variables I see the limitation of the results to the
compliers as one that was unintended, but driven by the lack of
identification for other subpopulations given the design of the study.
This limitation should be acknowledged, but one should not drop the
analysis simply because the original estimand cannot be identified.
Note that our case with instrumental variables is slightly worse than
in the gender example, because we cannot actually identify all
individuals with certainty as compliers.

There are alternatives to this view. One approach is to focus solely or
primarily on intention-to-treat effects. The strongest argument for
that is in the context of randomized experiments with noncompliance.
The causal interpretation of intention-to-treat effects is justified by
the randomization. As Freedman writes, ``Experimental data should
therefore be analyzed first by comparing rates or
averages, following the intention-to-treat principle. Such comparisons are
justified because the treatment and control groups are balanced, within the
limits of chance variation, by randomization'' (\cite*{Fre06}, page 701).
Even in that case one may wish to also report estimates of the local
average treatment effects because they may correspond more closely to
the object of ultimate interest. The argument for focusing on
intention-to-treat or reduced-form estimates is weaker in other
settings. For example, in the Fulton Fish Market demand and supply
application, the intention-to-treat effects are the effects of weather
conditions on prices and quantities. These effects may be of little
substantive interest to policy makers interested in tax policy. The
substantive interest for these policy makers is almost exclusively in
the \textit{structural} effects of price changes on demand and supply, and
reduced form effects are only of interest in sofar as they are
informative about those structural effects. Of course, one should bear
in mind that the reduced form or intention-to-treat effects rely on
fewer assumptions.

A second alternative is associated with the partial identification
approach by
Manski (\citeyear{Man90}, \citeyear{Man02}, \citeyear{Man03}, \citeyear{Man07}); % on what he calls a \textit{partial
%identification} approach.
see also \citet{Rob} and \citet{LeaN1} for antecedents. In this
setting that suggests maintaining the focus on the original estimand,
say the overall average effect, we cannot estimate that accurately
because we cannot estimate the average value of $Y_i(0)$ for
always-takers or the average value of $Y_i(1)$ for nevertakers, but we
can {bound} the average effect of interest because we know a priori  that the average value of $Y_i(0)$ for always-takers and the
average value of $Y_i(0)$ for nevertakers is restricted to lie in the
unit interval. Manski's is a principled and coherent approach. One
concern with the approach is that it has often focused on reporting
solely these bounds, leading researchers to miss relevant information
that is available given the maintained assumptions. Two different data
sets may lead to the same bounds even though in one case we may know
that the average effect for one subpopulation (the compliers) is
positive and statistically significantly different from zero whereas in
the other case there need not be any evidence of a nonzero effect for
any subpopulation. It would appear to be useful to distinguish between
such cases by reporting estimates of both the local average treatment
effect and the bounds.

%s5 #&#
\section{The Substantive Content of the Instrumental Variables Assumptions}
\label{section:content}

In this section, I will discuss the substantive content of the three
key assumptions, random assignment, the exclusion restriction and the
monotonicity assumption. I will not discuss here the fourth assumption,
instrument relevance. In practice, the main issue with that assumption
concerns the quality of inferences when the assumption is close to
being violated. See Section~\ref{section_weak} for more discussion, and \citet{StaSto97} for a detailed study.
%, and in the context of some examples, when they are plausible.

%s5.1 #&#
\subsection{Unconfoundedness of the Instrument}

First, consider the random assignment or unconfoundedness assumption.
In a slightly different setting, this is a very familiar assumption.
Matching methods often rely on random assignment, either
unconditionally or conditionally, for their justification.

In some of the leading applications of instrumental variables methods,
this assumption is satisfied by design, when the instrument is
physically randomized.
For example, in the draft lottery example (Angrist, \citeyear{Ang90}), draft
priority is used as an instrument for veteran status in an evaluation
of the causal effect of veteran status on mortality and earnings. In
that case, the instrument, the draft priority number was assigned by
randomization. Similarly, in the flu example (Hirano et al., \citeyear{Hiretal00}), the instrument for influenza vaccinations, the letter to
the physician, was randomly assigned.

In other cases, the conditional version of this assumption is more
plausible. In the \citet{McCNew94} study, proximity of an
individual to a hospital with particular facilities is used as an
instrument for the receipt of intensive
treatment of acute myocardial infarction. This proximity measure is not
randomly assigned, and McClellan and Newhouse use covariates to make
the unconfoundedness assumption more plausible. For example, they worry
about differences between individuals living in rural versus urban
areas. To adjust for such differences, they use as one of the
covariates the distance to the nearest hospital (regardless of the
facilities at the nearest hospital).

A key issue is that although on its own this random assignment or
unconfoundedness assumption justifies a causal interpretation of the
intention-to-treat effects, it is \textit{not} sufficient for a causal
interpretation of the instrumental variables estimand, the ratio of the
ITT effects for outcome and treatment.

%s5.2 #&#
\subsection{The Exclusion Restriction}
\label{section:exclusion}

Second, consider the exclusion restriction. This is the most critical
and typically most controversial assumption underlying instrumental
variables methods.

First of all, it has some testable implications; see \citet{BalPea97} and the recent discussions in
\citet{Kit09} and \citet{RamLau11}. This testable restriction can be seen most easily in
a binary outcome setting. Under the three assumptions, random
assignment, the exclusion restriction and monotonicity, the
intention-to-treatment effect of the assignment on the outcome is the
product of two causal effects. First, the average effect of the
assignment on the outcome for compliers, and second, the
intention-to-treat effect of the assignment on receipt of the
treatment, which is equal to the population proportion of compliers. If
the outcome is binary, the first factor is between $-1$ and 1. Hence, the
intention-to-treat effect of the assignment on the outcome has to be
bounded in absolute value by the intention-to-treat effect of the
assignment on the receipt of the treatment. This is a testable
restriction. If the outcomes are multivalued, there is in fact a range
of restrictions implied by the assumptions. However, there exist no
consistent tests that will reject the null hypothesis with probability
going to one as the sample size increases in all scenarios where the
null hypothesis is wrong.

Let us assess these restrictions in the flu example. Because
\begin{eqnarray*}
&& \operatorname{pr}(Y_i=1,X_i=0|Z_i=1)
\\
&& \quad =\operatorname{pr}\bigl(Y_i(0)=1|T_i=n\bigr)
\cdot \operatorname{pr}(T_i=n)
\end{eqnarray*}
and
\begin{eqnarray*}
&& \operatorname{pr}(Y_i=1,X_i=0|Z_i=0)
\\
&& \quad =\operatorname{pr}\bigl(Y_i(0)=1| T_i=n\mbox{ or } c\bigr)
\\
&& \qquad {}\cdot\operatorname{pr}(T_i=n \mbox{ or }c)
\\
&& \quad =\operatorname{pr}\bigl(Y_i(0)=1|T_i=n\bigr)
\cdot\operatorname{pr}(T_i=n) %{\mbox{pr}(\mbox{nevertaker or complier})}
\\
&& \qquad {}+\operatorname{pr}\bigl(Y_i(0)=1|T_i=c\bigr)
\cdot\operatorname{pr}(T_i=c) %{{\rm pr}({\rm nevertaker\ or\ complier})}
\end{eqnarray*}
it follows that
%
%e5.1 #&#
\begin{eqnarray}
\label{test} && \operatorname{pr}(Y_i=1,X_i=0|Z_i=1)
\nonumber
\\[-8pt]
\\[-8pt]
&& \quad \leq\operatorname{pr}(Y_i=1,X_i=0|Z_i=0).
\nonumber
\end{eqnarray}
There are three more restrictions in this setting with a binary
outcome, binary treatment and binary instrument; see \citet{ImbRub97N2}, \citet{BalPea97} and \citet{RicEvaRob11} for details.
For the flu data, the simple frequency estimator for the left-hand side
of (\ref{test}) is
$30/1389=0.0216$, and the right-hand side is $31/72=0.0211$, leading to
a slight violation as pointed out in \citet{RicEvaRob11} and Imbens and Rubin (\citeyear{ImbRub}). Although not statistically
significant, it shows that these restrictions have content in practice.

To assess the plausibility of the exclusion restriction, it is often
helpful to do so separately in subpopulations defined by compliance status.
Let us first consider the exclusion restriction for
always-takers,
who would receive
the influenza vaccine irrespective of the receipt of the
letter by their physician.
Presumably, such patients are generally at higher risk for the flu.
Why would such patients be
affected by a letter warning their physicians about the upcoming
flu season when they will get inoculated irrespective of this warning?
It may be that the letter led the physician to
take other actions beyond giving the flu vaccine,
such as encouraging the patient to avoid exposure.
These other actions may
affect health outcomes, in which case the exclusion restriction would
be violated.
The exclusion restriction for never-takers has different content.
These patients would not receive the vaccine in any case.
If their physicians
did not regard the risk of flu as sufficiently high to encourage their
patients to have the vaccination,
presumably the physician would not take other actions either. For these
patients, the exclusion restriction may therefore be reasonable.

Consider the draft lottery example. In that case, the always-takers are
individuals who volunteer for military service irrespective of their
draft priority number. It seems plausible that the draft priority
number has no causal effect on their outcomes. never-takers are
individuals who do not serve in the military irrespective of their
draft priority number. If this is for medical reasons, or more
generally reasons that make them ineligible to serve, this seems
plausible. If, on the other hand these are individuals fit but
unwilling to serve, they may have had to take actions to stay out of
the military that could have affected their subsequent civilian labor
market careers. Such actions may include extending their educational
career, or temporarily leaving the country.
Note that these issues are not addressed by the random assignment of
the instrument.

In general, the concern is that the instrument creates incentives not
only to receive the treatment, but also to take additional actions that
may affect the outcome of interest. The nature of these actions may
well differ by compliance type.
Most important is to keep in mind that this assumption is typically a
substantive assumption, not satisfied by design outside of
double-blind, single-dose placebo control randomized experiments with
noncompliance.

%s5.3 #&#
\subsection{Monotonicity}

Finally, consider the monotonicity or no-defiers assumption. Even
though this assumption is often the least controversial of the three
instrumental variables assumptions, it is still sometimes viewed with
suspicion. For example, whereas Robins views the assumption as ``often,
but not always reasonable'' (\cite*{Rob}, page 122), \citet{Fre06} wonders:
``The identifying restriction for the
instrumental-variables estimator
is troublesome: just why are there no defiers?''
(\cite*{Fre06}, page 700). In many applications, it is perfectly clear
why there should be no or at most few defiers.
The instrument plays the role of an \textit{incentive} for the individual
to choose the active treatment by either making it more attractive to
take the active treatment or less attractive to take the control
treatment. As long as individuals do not respond perversely to this
incentive, monotonicity is plausible with either no or a negligible
proportion of defiers in the population.
The term incentive is used broadly here: it may be a financial
incentive, or the provision of information, or an imperfectly monitored
legal requirement, but in all cases something that makes it more
likely, at the individual level, that the individual participates in
the treatment.

Let us consider some examples.
If noncompliance is one-sided, and those assigned to the control group
are effectively embargoed from receiving the treatment, monotonicity is
automatically satisfied. In that case $X_i(0)=0$, and there are no
always-takers or defiers. The example discussed in \citet{SomZeg91},
Imbens and Rubin (\citeyear{ImbRub97N1}) and \citet{Gre} fits this set up.

In the
flu application introduced in Section~\ref{section:modern}, the letter
to the physician creates an additional incentive for the physician to
provide the flu vaccine to a patient, something beyond any incentives
the physician may have had already to provide the vaccine.
Some individuals may already be committed to the vaccine, irrespective
of the letter (the always-takers), and some may not be swayed by the
receipt of the letter (the never-takers), and that is consistent with
this assumption. Monotonicity only requires that there is no patient,
who, if their physician receives the letter, would not take the
vaccine, whereas they would have taken the vaccine in the absence of
the letter.

Consider a second example, the influential draft lottery application by
\citet{Ang90} (see also \cite*{HeaNewHul}). Angrist is
interested in evaluating the effect of military service on subsequent
civilian earnings, using the draft priority established by the draft
lottery as an instrument. Monotonicity requires that assigning an
individual priority for the draft rather than not, may induce them to
serve in the military, or may not affect them, but cannot induce them
to switch from serving to not serving in the military.
Again that seems plausible. Having high priority for the draft
increases the cost of staying out of the military: that may not be
enough to change behavior, but it would be unusual if the increased
cost of staying out of the military induced an individual to switch
from serving in the military to not serving.

As a third example, consider
the \citet{PerHeb89} study of the effect of smoking on birthweight.
Permutt and Hebel use the random assignment to a smoking-cessation
program as an instrument for the amount of smoking.
In this case, the monotonicity assumption requires that there are no
individuals who as a causal effect of the assignment to the
smoking-cessation program end up smoking more. There may be individuals
who continue to smoke as much under either assignment and individuals
who reduce smoking as a result of the assignment, but the assumption is
that there is nobody who increases their smoking as a result of the
smoking-cessation program. In all these examples, monotonicity requires
individuals not to respond perversely to changes in incentives.
Systematic and major violations in such settings seem unlikely.

In other settings, the assumption is less attractive.
Suppose a program has assignment criteria that are checked by two
administrators. Individuals entering the assignment process are
assigned randomly to one of
the two administrators. The assignment criteria may be interpreted
slightly differently by the two administrators, with on average
administrator A being more strict than administrator B. Monotonicity
requires that anyone admitted by administrator A would also be admitted
by administrator B, or {vice-versa}. In this type of setting,
monotonicity does not appear to be as plausible as it is in the
settings where the instrument can be viewed as creating an incentive to
participate in the treatment.
For example, in an analysis of the effect of prison time on recidivism,
%Aiza (\citeyear{A10}) uses
Aizer and Doyle (\citeyear{AizDoy13}) use
random assignment of cases to judges, and in an
analysis of the effect of bankruptcy,
%Dobbie (\citeyear{D11}) uses
Dobbie and Song (\citeyear{DobSon13}) use
random
assignment of bankruptcy applications to judges.

The discussion in this section focuses primarily on the case with a
binary treatment and a binary instrument. In cases with multivalued
treatments, the monotonicity can be generalized in two different ways.
In both cases, it may be less plausible than in the binary case. Let
$X_i(z)$ be the potential treatment level associated with the
assignment $z$. One can generalize the monotonicity assumption for the
binary instrument case to this case as
%
%e5.2 #&#
\begin{eqnarray}
&& X_i(z) \mbox{ is nondecreasing in } z \quad \mbox{for all } i
\nonumber\\
\eqntext{\mbox{(monotonicity in instrument)}.}
\end{eqnarray}
This generalization is used in \citet{AngImb95}. It is
consistent with the view of the instrument as changing the incentive to
participate in the treatment: increasing the incentive cannot decrease
the level of the treatment received.
Angrist and Imbens show that this assumption has testable implications.

An alternative generalization is
%
%e5.3 #&#
\begin{eqnarray}
&& \mbox{if } X_i(z)> X_j(z)\quad
\nonumber\\
&& \quad \mbox{then } X_i\bigl(z'\bigr)\geq
X_j\bigl(z'\bigr) \quad \mbox{for all }
z,z',i,j
\nonumber\\
\eqntext{\mbox{(monotonicity in unobservables).}}
\end{eqnarray}
This assumption, referred to as \textit{rank preservation} in \citet{Rob86}, implicitly ranks all units in terms of some unobservables
(\cite{ImbN1}). It assumes this ranking is invariant to the level of
the instrument. It implies that if $X_i(z)>X_j(z)$, then it cannot be
that $X_j(z')>X_i(z')$. It is equivalent to the ``continuous
prescribing preference'' in \citet{HerRob06}.

In both cases, the special case with a binary treatment is identical to
the previously stated monotonicity. In settings with multivalued
treatments, these assumptions are more restrictive than in the binary
treatment case. In the demand and supply example in Section~\ref{section:supplydemand} with linear supply and demand functions, both
the monotonicity in the instrument and monotonicity in the
unobservables conditions are satisfied.

%s6 #&#
\section{The Link to the Textbook Discussions of Instrumental Variables}
\label{section:textbook}

Most textbook discussions of instrumental variables use a framework
that is quite different at first sight from the potential outcome set
up used in Sections~\ref{section:modern} and~\ref{section:content}. These textbook discussions
(graduate texts include Wooldridge, \citeyear{Woo10}; \cite*{AngPis09}; \cite*{Gre11}; and \cite*{Hay00}, and introductory undergraduate
textbooks include \cite*{Woo08}; and \cite*{StoWat10})
are often closer to the simultaneous equations example from Section~\ref{section:supplydemand}. An exception is \citet{Man07} who uses the
potential outcome set up used in this discussion. In this section I
will discuss the standard textbook set up and relate it to the
potential outcome framework and the simultaneous equations set up.

The textbook version of instrumental variables does not explicitly
define the potential outcomes. Instead the starting point is a linear
regression function describing the relation between the realized
(observed) outcome $Y_i$, the endogenous regressor of interest $X_i$
and other regressors $V_i$:
%
%e6.1 #&#
\begin{equation}
\label{eq:classical} Y^\obs_i=\beta_0+
\beta_1 X_i+\beta_2'V_i+
\varepsilon_i.
\end{equation}
These other regressors a well as the instruments are often referred to
in the econometric literature as \textit{exogenous} variables. Although
this term does not have a well-defined meaning, informally it includes
variables that \citet{Cox92} called \textit{attributes}, as well as potential
causes whose assignment is unconfounded.
This set up covers both the demand function setting and the randomized
experiment example.
Although this equation looks like a standard regression function, that
similarity is misleading. Equation (\ref{eq:classical}) is not an
ordinary regression function in the sense that the first part does \textit{not} represent the conditional expectation of the outcome $Y_i$ given
the right-hand side variables $X_i$ and $V_i$.
Instead it is what is sometimes called a \textit{structural equation}
representing the causal response to changes in the input~$X_i$.

The key assumption in this formulation is that the unobserved component
$\varepsilon_i$ in this regression function is independent of the
exogenous regressors $V_i$ and the
instruments $Z_i$, or, formally
%
%e6.2 #&#
\begin{equation}
\label{eq:class_ass} \varepsilon_i \perp ( Z_i,V_i
).
\end{equation}
The unobserved component is \textit{not} independent of the endogenous
regressor $X_i$ though.
The value of the regressor $X_i$ may be partly chosen by individual $i$
to optimize some objection function as in the noncompliance example, or
the result of an equilibrium condition as in the supply and demand
model. The precise relation between $X_i$ and $\varepsilon_i$ is often
not fully specified.

How does this set up relate to the earlier discussion involving
potential outcomes?
Implicitly, there is in the background of this set up a causal,
unit-level response function. In the potential outcome notation, let
$Y_i(x)$ denote this causal response function for unit $i$, describing
for each value of $x$ the potential outcome corresponding to that level
of the treatment for that unit. Suppose the conditional expectation of
this causal response function is linear in $x$ and some exogenous covariates:
%
%e6.3 #&#
\begin{equation}
\label{lineareq} \mathbb{E}\bigl[Y_i(x)|V_i\bigr]=\beta
_0+\beta_1\cdot x+\beta_2^{\prime}V_i.
\end{equation}
Moreover, let us make the (strong) assumption that the difference
between the response function $Y_i(x)$ and its conditional expectation
does not depend on $x$, so we can define
the residual unambiguously as
\[
\varepsilon_i=Y_i(x)- \bigl(\beta_0+
\beta_1\cdot x+\beta _2'V_i
\bigr),
\]
with the equality holding for all $x$.
The residual $\varepsilon_i$ is now uncorrelated with $V_i$ by
definition. We will assume that it is in fact independent of $V_i$. Now
suppose we have an instrument $Z_i$ such that
\[
Y_i(x)\perp Z_i | V_i.
\]
This assumption is, given the linear representation for $Y_i(x)$,
equivalent to
\[
\varepsilon_i\perp Z_i | V_i.
\]
In combination with the assumption that $\varepsilon_i\perp V_i$, this
gives us the textbook version of the assumption given in (\ref{eq:class_ass}).
We observe $V_i$, $X_i$, the instrument $Z_i$, and the realized outcome
\[
Y^\obs_i=Y_i(X_i)=
\beta_0+\beta_1 X_i+\beta_2'V_i+
\varepsilon_i,
\]
which is the starting point in the econometric textbook discussion
(\ref{eq:classical}).

This set up is more restrictive than it needs to be. For example, the
assumption that the
difference between the response function $Y_i(x)$ and its conditional
expectation does not depend on $x$ can be relaxed to allow for
variation in the slope coefficient,
\[
Y_i(x)-Y_i(0)=\beta_1\cdot x+
\eta_i\cdot x,
\]
as long as the $\eta_i$ satisfies conditions similar to those on
$\varepsilon_i$.
The modern literature (e.g., \cite*{Mat07}) discusses such models
in more detail.

One key feature of the textbook version is that there is no separate
role for the monotonicity assumption. Because the linear model
implicitly assumes that the per-unit causal effect is constant across
units and levels of the treatment, violations of the monotonicity
assumption do not affect the interpretation of the estimand.
A second feature of the textbook version is that the exclusion
restriction and the random assignment assumption are combined in (\ref
{eq:class_ass}). Implicitly, the exclusion restriction is captured by
the absence of $Z_i$ in the equation (\ref{eq:classical}), and the
(conditional) random assignment is captured by (\ref{eq:class_ass}).

%s7 #&#
\section{Extensions and Generalizations}\label{section:extensions}

In this section, I will briefly review some of other approaches taken
in the instrumental variables literature. Some of these originate in
the statistics literature, some in the econometrics literature. They
reflect different concerns with the traditional instrumental variables
methods, sometimes because of different applications, sometimes because
of different traditions in econometrics and statistics. This discussion
is not exhaustive. I~will focus on highlighting the most interesting
developments and provide some references to the relevant literature.

%s7.1 #&#
\subsection{Model-based Approaches to Estimation and Inference}

Traditionally, instrumental variables analyses relied on linear
regression methods. Additional explanatory variables are incorporated
linearly in the regression function. The recent work in the statistics
literature has explored more flexible approaches to include covariates.
These approaches often involve modeling the conditional distribution of
the endogenous regressor given the instruments and the exogenous
variables. This is in contrast to the traditional econometric
literature which has focused on settings and methods that do not rely
on such models.

Robins (\citeyear{Rob}, \citeyear{Rob94}), \citet{HerRob06}, \citet{Gre},
Robins and Rotnitzky (\citeyear{RobRot04}) and \citet{Tan10} developed an approach that
allow for identification of average treatment effect by adding
parametric modelling assumptions. This approach starts with the
specification of what they call the \textit{structural mean}, the
expectation of $Y_i(x)$. This structural mean can be the conditional
mean given covariates, or the marginal mean, labeled the \textit{marginal
structural mean}. The specification for this expectation is typically
parametric. Then estimating equations for the parameters of these
models are developed. In the simple setting considered here, this would
typically lead to the same estimators considered already. An important
virtue of the method is that it has been extended to much more general
settings, in particular with time-varying covariates and dynamic
treatment regimes in a series of papers. In other settings, it has also
led to the development of doubly robust estimators (\cite{RobRot04}).
A~key feature of the models is that the models are robust in a
particular sense. Specifically, the estimators for the average
treatment effects are consistent irrespective of the misspecification
of the model, in the absence of intention-to-treat effects (what they
call the conditional ITT null).

\citet{ImbRub97N1} and
Hirano et al. (\citeyear{Hiretal00}) propose building a parametric
model for the compliance status in terms of additional covariates,
combined with models for the potential outcomes conditional on
compliance status and covariates.
Given the monotonicity assumption, there are three compliance types:
never-takers, always-takers and compliers. A natural model for
compliance status given individual characteristics $V_i$ is therefore a
trinomial logit model:
\begin{eqnarray*}
\pr(T_i=n|V_i=v) &=& \frac{\exp(v'\gamma_n)}{1+\exp(v'\gamma_n)+\exp
(v'\gamma_n)},
\\
\pr(T_i=a|V_i=v) &=& \frac{\exp(v'\gamma_a)}{1+\exp(v'\gamma_n)+\exp
(v'\gamma_n)}
\end{eqnarray*}
and
\[
\pr(T_i=c|V_i=v)=\frac{1}{1+\exp(v'\gamma_n)+\exp(v'\gamma_n)}.
\]
With continuous outcomes, the conditional outcome distributions given
compliance status and covariates may be normal:
\[
Y_i(x)|T_i=t, \quad V_i=v\sim \mathcal{N}
\bigl(\beta_{tx}' v,\sigma^2_{tx}
\bigr),
\]
for $(t,x)=(n,0),(a,1),(c,0),(c,1)$.
With binary outcomes, one may wish to use logistic regression models here.
This specification defines the likelihood function.
Hirano et al. (\citeyear{Hiretal00}) apply this to the flu data
discussed before.
Simulations in Richardson, Evans and Robins (\citeyear{RicEvaRob11}) suggest that the
modeling of the compliance status here is key. Specifically, they point
out that even in the absence of ITT effects there can be biases if the
model of the compliance status is misspecified.

Like Hirano et al. (\citeyear{Hiretal00}), \citet{RicEvaRob11} build parametric model only for the identified
distributions. They use them to estimate the bounds so that the
parametric assumptions do not contain identifying information.

%Little and Yau (\citeyear{L01}) similarly
Little and Yau (\citeyear{LitYau98}) and Yau and Little (\citeyear{YauLit01})
similarly
model the conditional expectation of
the outcome given compliance status and covariates. In their
application, there are no always-takers, only never-takers and
compliers. Their specification specifies parametric forms for the
conditional means given the compliance types and the treatment status:
\begin{eqnarray*}
\mathbb{E}\bigl[Y_i(0)|T_i=n,V_i=v\bigr] &=&
\beta_{n0}+\beta_{n1}'v,
\\
\mathbb{E}\bigl[Y_i(0)|T_i=c,V_i=v\bigr] &=&
\beta_{c00}+\beta_{c01}'v
\end{eqnarray*}
and
\[
\mathbb{E}\bigl[Y_i(1)|T_i=c,V_i=v\bigr]=
\beta_{c00}+\beta_{c11}'v.
\]

%s7.2 #&#
\subsection{Principal Stratification}\label{sec72}

\citet{FraRub02} generalize the latent compliance type
approach to instrumental variables in an important and novel way. Their
focus is on the causal effect of a binary treatment on some outcome.
However, it is not the average effect of the treatment they are
interested in, but the average within a subpopulation.
It is the way this subpopulation is defined that creates the
complications as well as the connection to instrumental variables.
There is a post-treatment variable that may be affected by the
treatment. Frangakis and Rubin postulate the existence of a pair of
potential outcomes for this post-treatment variable. The subpopulation
of interest is then defined by the values for the pair of potential
outcomes for this post-treatment variables.

Let us consider two examples: first, the randomized experiment with
noncompliance. The treatment here is the random assignment. The
post-treatment variable is the actual receipt of the treatment. The
pair of potential outcomes for this post-treatment variable captures
the compliance status. The subpopulation of interest is the
subpopulation of compliers.

The second example shows how principal stratification generalizes the
instrumental variables set up to other cases.
Examples of this type are considered in \citet{ZhaRubMea09},
Frumento et al. (\citeyear{Fruetal12}) and \citet{Rob86}.
Suppose we have a randomized experiment with perfect compliance. The
primary outcome is survival after one year. For patients who survive, a
quality of life measure is observed. We may be interested in the effect
of the treatment on quality of life. This is only defined for patients
who survive up to one year. The principal stratification approach
suggests focusing on the subpopulation or \textit{principal stratum} of
patients who survive irrespective of the treatment assignment.
Membership in this stratum is not observed, and so we cannot directly
estimate the average effect of the treatment on quality of life for
individuals in this stratum, but the data are generally still
informative about such effects, particularly under monotonicity assumptions.

%s7.3 #&#
\subsection{Randomization Inference with Instrumental~Variables}

Most of the work on inference in instrumental variables settings is
model-based. After specifying a model relating the treatment to the
outcome, the conditional distribution or conditional mean of outcomes
given instruments is derived. The resulting inferences are conditional
on the values of the instruments. A very different approach is taken in
\citet{Ros96} and Imbens and Rosenbaum (\citeyear{ImbRos05}).

Rosenbaum focuses on the distribution for statistics generated by the
random assignment of the instruments. In the spirit of the work by
\citet{Fis25} confidence intervals for the parameter of interest,
$\beta_1$ in equation (\ref{lineareq}) are based on this randomization
distribution.
Similar to confidence intervals for treatment effects based on
inverting conventional Fisher $p$-values, these intervals have exact
coverage under the stated assumptions.
However, these results rely on arguably restrictive constant treatment
effect assumptions.

%s7.4 #&#
\subsection{Matching and Instrumental Variables}
\label{section_matchinginstruments}

In many observational studies using instrumental variables approaches,
the instruments are not randomly assigned. In that case, adjustment for
additional pretreatment variables can sometimes make causal inferences
more credible. Even if the instrument is \mbox{randomly} assigned, such
adjustments can make the inferences more precise. Traditionally, in
econometrics these adjustments are based on regression methods.
Recently, in the statistics literature matching methods have been
proposed as a way to do the adjustment for pretreatment variables
(Baiocchi et al., \citeyear{Baietal10}).

%s7.5 #&#
\subsection{Weak Instruments}
\label{section_weak}

One concern that has arisen in the econometrics literature is about
\textit{weak} instruments. For an instrument to be helpful in estimating
the effect of the treatment, it not only needs to have no direct effect
on the outcome, it also needs to be correlated with the treatment.
Suppose this correlation is very close to zero. In the simple case, the
IV estimator is the ratio of covariances,
\begin{eqnarray*}
\hat\beta_{1,\iv} &=& \frac{\widehat{\operatorname{cov}}(Y_i,Z_i)}{\widehat{\operatorname{cov}}(X_i,Z_i)} \\
& = & \frac{({1}/{N})\sum_{i=1}^N (Y_i-\overline{Y}) (Z_i-\overline{Z})}{
({1}/{N})\sum_{i=1}^N (X_i-\overline{X}) (Z_i-\overline{Z})}.
\end{eqnarray*}
The distribution of this ratio can be approximated by a normal
distribution in large samples, as long as the covariance in the
denominator is nonzero in the population.
If the population value of the covariance in the denominator is exactly
zero, the distribution of the ratio $\hat\beta_{1,\iv}$ is Cauchy in
large samples, rather than normal (\cite{Phi89}; \cite{StaSto97}). The weak instrument literature is concerned with the
construction of confidence intervals in the case the covariance is
close to zero. Interest in this problem rose sharply after a study by
\citet{AngKru91}, which remains the primary empirical
motivation for this literature. Angrist and Krueger were interested in
estimating the causal effect of years of education on \mbox{earnings}. They
exploited variation in educational achievement by quarter of birth
attributed to differences in compulsory schooling laws. These
differences in average years of education by quarter of birth were
small, and they attempted to improve precision of their estimators by
including interactions of the basic instruments, the three quarter of
birth dummies, with indicators for year and state of birth. \citet{BouJaeBak95} showed that the estimates using the
interactions as additional instruments were potentially severely
affected by the weakness of the instruments. In one striking analysis,
they reestimated the Angrist--Krueger regressions using randomly
generated quarter of birth data (uncorrelated with earnings or years of
education). One might have expected, and hoped, that in that case one
would find an imprecisely estimated effect. Surprisingly, \citet{BouJaeBak95} found that the confidence intervals constructed by
Angrist and Krueger suggested precisely estimated effects for the
effect of years of education on earnings. It was subsequently found
that with weak instruments the TSLS estimator, especially with many
instruments, was biased, and that the standard variance estimator led
to confidence intervals with substantial undercoverage (\cite{BouJaeBak95}; \cite{StaSto97}; \cite{ChaImb04}).

Motivated by the Bound--Jaeger--Baker findings, the weak and many
instruments literature focused on point and interval estimators
with better properties in settings with weak instruments.
Starting with
\citet{StaSto97}, a literature developed to construct
confidence intervals for the instrumental variables estimand that
remained valid irrespective of the strength of the instruments. A key
insight was that confidence intervals based on the inversion of
Anderson--Rubin (\citeyear{A49}) statistics have good properties in settings with
weak instruments; see also \citet{Mor03}, \citet{AndSto},
\citet{Kle02} and \citet{AndMorSto06}.

Let us look at the simplest case  with a single endogenous
regressor, a single instrument, and no additional regressors and
normally distributed residuals:
\[
Y_i(x)=\beta_0+\beta_1\cdot x+
\varepsilon_i\quad \mbox{with } \varepsilon_i
|Z_i\sim\mathcal{N} \bigl(0,\sigma^2_\varepsilon
\bigr).\vadjust{\goodbreak}
\]
The Anderson--Rubin statistic
is, for a given value of $b$
\begin{eqnarray*}
\operatorname{AR}(b) &=&  \Biggl(\frac{1}{\sqrt N}\sum_{i=1}^N
(Z_i-\oz)\cdot (Y_i-b\cdot X_i)
\Biggr)^2 \\
&&{}\bigg/ \Biggl(\frac{1}{N}\sum
_{i=1}^N (Z_i-\oz)^2 \cdot
\hat\sigma ^2_\varepsilon \Biggr),
\end{eqnarray*}
where $\oz=\sum_{i=1}^N Z_i/N$, and for some estimate of the residual
variance $\sigma^2_\varepsilon$. At the true value $b=\beta_1$, the
AR statistic has in large samples a chi-squared distribution with one
degree of freedom. \citet{StaSto97} propose constructing a
confidence interval by inverting this test statistic:
\[
\mathrm{CI}^{0.95}(\beta_1)= \bigl\{b| \operatorname{AR}(b)\leq3.84 \bigr\}.
\]
The subsequent literature has extended this by allowing for multiple
instruments and developed various alternatives, all with the focus on
methods that remain valid irrespective of the strength of the
instruments; see \citet{AndSto} for an overview of this literature.

%s7.6 #&#
\subsection{Many Instruments}
\label{section_many}

Another strand of the literature motivated by the Angrist--Krueger
study focused on settings with many weak instruments. The concern
centered on the  Bound, Jaeger and Baker (\citeyear{BouJaeBak95})  finding that in a setting
similar to the Angrist--Krueger setting using TSLS with many randomly
generated instruments led to confidence intervals that had very low
coverage rates.

To analyze this setting,
Bekker (\citeyear{Bek94}) considered the behavior of various estimators under an
asymptotic sequence where the number of instruments increases with the
sample size.
Asymptotic approximations to sampling distributions based on this
sequence turned out to be much more accurate than those based on
conventional asymptotic approximations.
A key finding in Bekker (\citeyear{Bek94}) is that under such sequences one of the
leading estimators, Two-Stage-Least-Squares (TSLS, see the \hyperref[appA]{Appendix} for
details) estimator is no longer consistent, whereas another estimator,
Limited Information Maximum Likelihood (LIML, again see the \hyperref[appA]{Appendix}
for details) estimator remains consistent although the variance under
this asymptotic sequence differs from that under the standard sequence;
see also
Kunitomo (\citeyear{K80}), Morimune (\citeyear{M83}),
Bekker and van der Ploeg (\citeyear{B05}),
\citet{ChaImb04}, \citet{ChaSwa05}, Hahn (\citeyear{H02}),
Hansen, Hausman and Newey (\citeyear{H08}),
Koles\'ar et al. (\citeyear{K13}).%and Van Hasselt (\citeyear{H10}).

%s7.7 #&#
\subsection{Proxies for Instruments}

\citet{HerRob06} and
\citet{Cha11} explores settings where the instrument is not directly
observed. Instead a proxy variable $Z^*_i$ is observed. This proxy
variable is correlated with the underlying instrument $Z_i$, but not
perfectly so.
The potential outcomes $Y_i(z,x)$ are still defined in terms of the
underlying, unobserved instrument $Z_i$.
The unobserved \mbox{instrument} $Z_i$ satisfies the instrumental  variables
assumptions, random assignment, the exclusion restriction and the
monotonicity assumption. In addition, the observed proxy $Z_i^*$ satisfies
\begin{eqnarray*}
Z_i^* &\perp &  Y_i(0,0),Y_i(0,1),Y_i(1,0),\\
&& {} Y_i(1,1),X_i(0),X_i(1)|Z_i.
\end{eqnarray*}
Chalak shows that the ratio of covariances (now no longer the ratio of
intention-to-treat effects) still has an
interpretation
%interpretion
of an average
causal effect.

%s7.8 #&#
\subsection{Regression Discontinuity Designs}

Regression Discontinuity (RD) designs attempt to estimate causal
effects of a binary treatment in settings where the assignment
mechanism is a deterministic function of a pretreatment variable.
In the sharp version of the RD design, the assignment mechanism takes
the form
\[
X_i=\mathbf{1}_{V_i\geq c},
\]
for some fixed threshold $c$: all units with a value for the covariate
$V_i$ exceeding $c$ receive the treatment and all units with a value
for $V_i$ less than $c$ are in the control group. Under smoothness
assumptions, it is possible in such settings to estimate the average
effect of the treatment for units with a value for the pretreatment
variable equal to $V_i\approx c$:
\begin{eqnarray*}
&& \mathbb{E}\bigl[Y_i(1)-Y_i(0)|V_i=c
\bigr] \\
&& \quad =\lim_{w\uparrow c} \mathbb {E}[Y_i|V_i=w]-
\lim_{w\downarrow c} \mathbb{E}[Y_i|V_i=w].
\end{eqnarray*}
These designs were introduced by
\citet{ThiCam60}, and have been used in psychology,
sociology, political science and economics. For example, many
educational programs have eligibility criteria that allow for the
application of RD methods;
see \citet{Coo08} for a recent historical perspective and \citet{ImbWoo09} for a recent review.

A generalization of the sharp RD design is the \textit{Fuzzy Regression
Discontinuity} or FRD design. In this case, the probability of receipt
of the treatment increases discontinuously at the threshold,
but not necessarily from zero to one:
\[
\lim_{w\downarrow c}\pr(X_i=1|V_i=w)\neq\lim
_{w\uparrow c}\pr(X_i=1|V_i=w).
\]
In that case, it is no longer possible to consistently estimate the
average effect of the treatment for all units at the threshold.
Hahn, Todd and Van der Klaauw (\citeyear{HahTodvan00}) demonstrate that there is a close
link to the instrumental variables set up. Specifically Hahn, Todd and
Van der Klaauw show that one can estimate a local average treatment
effect at the threshold. To be precise, one can identify the average
effect of the treatment for those who are on the margin of getting the
treatment:
\begin{eqnarray*}
&& \mathbb{E}\Bigl[Y_i(1)-Y_i(0) \big|\\
&& \quad\hspace*{2pt} {}V_i=c,\lim
_{w\uparrow c} X_i(w)=0, \lim_{w\downarrow c}
X_i(w)=1 \Bigr]\\
&& \quad =
\frac{\lim_{w\uparrow c} \mathbb{E}[Y_i|V_i=w]-\lim_{w\downarrow c} \mathbb{E}[Y_i|V_i=w]}{
\lim_{w\uparrow c} \mathbb{E}[X_i|V_i=w]-\lim_{w\downarrow c}
\mathbb{E}[X_i|V_i=w]}.
\end{eqnarray*}
This estimand can be estimated as the ratio of an estimator for the
discontinuity in the regression function for the outcome and an
estimator for the discontinuity in the regression function for the
treatment of interest.

%s8 #&#
\section{Conclusion}
\label{section:conclusion}

In this paper, I review the connection between the recent statistics
literature on instrumental variables and the older econometrics
literature. Although the econometric literature on instrumental
variables goes back to the 1920s, until recently it had not made much
of an impact on the statistics literature. The recent statistics
literature has combined some of the older insights from the
econometrics instrumental variables literature with the separate
literature on causality, enriching both in the process.

%sA #&#
\begin{appendix}
\section*{Appendix: Estimation and Inference,
Two-Stage-Least-Squares and Other
Traditional Methods}\label{appA}

%sA.1 #&#
\subsection{Set up}

In this section, I will discuss the traditional econometric approaches
to estimation and inference in instrumental variables settings. Part of
the aim of this section is to provide easier access to the econometric
literature and terminology on instrumental variables, and to provide a
perspective and context for the recent advances.

The textbook setting is the one discussed in the previous section,
where a scalar outcome $Y_i$ is linearly related to a scalar covariate
of interest
$X_i$. In addition, there may be additional exogenous covariates $V_i$.
The traditional model is
%
%eA.1 #&#
\begin{equation}
\label{eq:outcome}
Y_i=\beta_0+\beta_1
X_i+\beta_2'V_i+
\varepsilon_i.
\end{equation}
In addition, we have a vector of instrumental variables $Z_i$, with
dimension $K$.

An important distinction in the traditional econometric literature is
between the case with a single instrument ($K=1$), and the case with
more than one instrument ($K>1$). More generally, with more than one
endogenous regressor, the distinction is between the case with the
number of instruments equal to the number of endogenous regressors and
the case with the number of instruments larger than the number of
endogenous regressors. In the empirical literature, there are few
credible examples with more than one endogenous regressor, so I focus
here on the case with a single endogenous regressor.
The first case, with a single instrument, is referred to as the \textit{just-identified} case, and the second, with multiple instruments and a
single endogenous regressor, as the \textit{over-identified} case. In the
textbook setting with a linear model and constant coefficients, this
distinction has motivated different estimators and specification tests.
In the modern literature, with its explicit allowance for heterogeneity
in the treatment effects, these tests, and the distinction between the
various estimators, are of less interest. In the recent statistics
literature, little attention has been paid to the over-identified case
with multiple instruments. An exception is \citet{Sma07}.

Obviously, it is often difficult in applications to find even a single
variable that satisfies the conditions for it to be a valid instrument.
This raises the question how relevant the literature focusing on
methods to deal with multiple instruments is for empirical practice.
There are two classes of applications where multiple instruments could
credible arise. First, suppose one has a single continuous (or
multivalued) instrument that satisfies the instrumental variables
assumptions, monotonicity, random assignment and the exclusion
restriction. Then any monotone function of the instruments also
satisfies these assumptions, and one can use multiple monotone
functions of the original instrument as instruments.
Second,
if one has a single instrument in combination with exogenous
covariates, then one can use interactions of the instrument and the
covariates to generate additional instruments.

Consider, for example, the Fulton fish market study by Graddy (\citeyear{Gra95},
\citeyear{Gra96}). Graddy uses weather conditions as an instrument that affects
supply but not demand. Specifically, she measures wind speed and wave
height, giving her two basic instruments. She also constructs functions
of these basic instruments, such as indicators that the wind speed or
wave height exceeds some threshold.

%sA.2 #&#
\subsection{The Just-Identified Case with no Additional~Covariates}

The traditional approach to estimation in this case is to use what is
known in the econometrics literature as the instrumental variables
estimator. In the case without additional exogenous covariates, the
most widely used estimator is simply the ratio of two covariances:
\begin{eqnarray*}
\hat{\beta}_{1,\iv} &=& \frac{\widehat{\operatorname{cov}}(Y_i,Z_i)}{\widehat{\operatorname{cov}}(X_i,Z_i)} \\
&=& \frac{({1}/{N})\sum_{i=1}^N (Y_i-\overline{Y}) (Z_i-\overline{Z})}{
({1}/{N})\sum_{i=1}^N (X_i-\overline{X}) (Z_i-\overline{Z})},
\end{eqnarray*}
where $\overline{Y}$, $\overline{Z}$ and $\overline{X}$ denote
sample averages.
If the instrument $Z_i$ is binary, this is also known as the Wald estimator:
\[
\hat\beta_{1,\iv}=\frac{\oy_1-\oy_0}{
\ox_1-\ox_0},
\]
where for $z=0,1$
\[
\oy_z=\frac{1}{N_z}\sum_{i:Z_i=z}
Y_i,\quad \ox_z=\frac{1}{N_z}\sum
_{i:Z_i=z} X_i,%\hskip0.5cm
% \ox_1=\frac{1}{N_1}\sum_{i:Z_i=1} X_i,
\]
and $N_1=\sum_{i=1}^N Z_i$ and $N_0=\sum_{i=1}^N (1-Z_i)$.

One can interpret this estimator in two different ways. These
interpretations are useful for motivating extensions to settings with
multiple instruments and additional exogenous regressors. First, the
\textit{indirect least squares} interpretation. This relies on first
estimating separately the two \textit{reduced form regressions}, the
regressions of the outcome on the instrument:
\[
Y_i=\pi_{10}+\pi_{11}\cdot Z_i+
\varepsilon_{1i},
\]
and the regression of the endogenous regressor
on the instrument:
\[
X_i=\pi_{20}+\pi_{21}\cdot Z_i+
\varepsilon_{2i}.
\]
The indirect least squares estimator is the ratio of the least squares
estimates of $\pi_{11}$ and $\pi_{21}$, or $\hat\beta_{1,\mathrm{ils}}=\hat\pi_{11}/\hat\pi_{21}$. Note that in the randomized
experiment example where $X_i$ and $Z_i$ are binary, the $\pi_{11}$
and $\pi_{12}$ are the \textit{intention-to-treat} effects,
with
$\hat\pi_{11}=\oy_1-\oy_0$ and
$\hat\pi_{12}=\ox_1-\ox_0$.

Second, I discuss the two-stage-least-squares interpretation of the
instrumental variables estimator. First, estimate the
reduced form regression of the treatment on the instruments and the
exogenous covariates. Calculate the predicted value for the endogenous
regressor from this regression:
\[
\hat X_i=\hat\pi_{20}+\hat\pi_{21}\cdot
Z_i.
\]
The estimate the regression of the outcome on the predicted endogenous
regressor and the additional covariates,
\[
Y_i=\beta_0+\beta_1 \hat
X_i+\eta_i,
\]
by least squares to get the TSLS estimator $\hat\beta_{\tsls}$.
In this just-identified setting, the three estimators for $\beta_1$
are numerically identical:
$\hat\beta_{1,\iv}=\hat\beta_{1,\ils}=\hat\beta_{1,\tsls}$.

%sA.3 #&#
\subsection{The Just-Identified Case with Additional~Covariates}

In most econometric applications, the instrument is not physically
randomized. There is in those cases no guarantee that the instrument is
independent of the potential outcomes. Often researchers use covariates
to weaken the requirement on the instrument to conditional independence
given the exogenous covariates.
In addition, the additional exogenous covariates can serve to increase
precision.
In that case with additional covariates, the estimation strategy
changes slightly.
The two reduced form regressions now take the form
\[
Y_i=\pi_{10}+\pi_{11}\cdot Z_i+
\pi_{12}' V_i+\varepsilon_{1i},
\]
and the regression of the endogenous regressor
on the instrument:
\[
X_i=\pi_{20}+\pi_{21}\cdot Z_i+
\pi_{22}' V_i+\varepsilon_{2i}.
\]
The indirect least squares estimator is again the ratio of the least
squares estimates of $\pi_{11}$ and $\pi_{21}$, or $\hat\beta
_{1,\mathrm{ils}}=\hat\pi_{11}/\hat\pi_{21}$.

For the two-stage-least-squares estimator, we again first estimate the
regression of the endogenous regressor on the instrument, now also
including the exogenous regressors. The next step is to predict the
endogenous covariate:
\[
\hat{X}_i=\hat\pi_{20}+\hat\pi_{21}\cdot
Z_i+\hat\pi_{22}' V_i .
\]
Finally, the outcome is regressed on the predicted value of the
endogenous regressor and the actual values of the exogenous variables:
\[
Y_i=\beta_0+\beta_1 \hat
X_i+\beta_2'V_i+
\eta_i.
\]
The TSLS estimator is again identical to the ILS estimator.

For inference, the traditional approach is to assume homoscedasticity
of the residuals $Y_i-\beta_0-\beta_1 X_i-\beta_2'V_i$ with variance
$\sigma^2_\varepsilon$.
In large samples, the distribution of the estimator $\hat{\beta}_{\mathrm{iv}}$ is approximately normal, centered around the true value $\beta
_1$. Typically, the variance is estimated as
\[
\widehat{\mathbb{V}}=\hat\sigma^2_\varepsilon\cdot \left( \left(
\begin{array} {@{}c@{}} 1
\\
\hat X_i
\\
V_i \end{array} %
 \right) \left( %
\begin{array} {@{}c@{}} 1
\\
\hat X_i
\\
V_i \end{array} %
 \right)^{\prime} \right)^{-1} .
\]
See the textbook discussion in Wooldridge (\citeyear{Woo10}).

%sA.4 #&#
\subsection{The Over-Identified Case}

The second case of interest is the overidentified case. The main
equation remains
\[
Y_i=\beta_0+\beta_1 X_i+
\beta_2'V_i+\varepsilon_i,
\]
but now the instrument $Z_i$ has dimension $K>1$.
We continue to assume that the residuals $\varepsilon_i$ are
independent of the instruments with mean zero and variance $\sigma
^2_\varepsilon$.
This case is the subject of a large literature, and
many estimators have been proposed. I will briefly discuss two. For a
more detailed discussion, see Wooldridge (\citeyear{Woo10}).

%sA.5 #&#
\subsection{Two-Stage-Least-Squares}

The TSLS approach extends naturally to the setting with multiple instruments.
First, estimate the reduced form regression of the endogenous variable
$X_i$ on the instruments $Z_i$ and the exogenous variables $V_i$,
\[
X_i=\pi_{20}+\pi_{21}'Z_i+
\pi_{22}'V_i+\varepsilon_{2i},
\]
by least squares. Next, calculate the predicted value,
\[
\hat{X}_i=\hat{\pi}_{20}+\hat{\pi}_{21}'Z_i+
\hat{\pi}_{22}'V_i .
\]
Finally, regress the outcome on the predicted value from this regression:
\[
Y_i=\beta_0+\beta_1 \hat
X_i+\beta_2'V_i+
\eta_i.
\]
The fact that the dimension of the instrument $Z_i$ is greater than one
does not affect the mechanics of the procedure.

To illustrate this, consider the Graddy Fulton Fish Market data.
Instead of simply using the binary indicator stormy/not-stormy as the
instrument, we can use the trivalued weather indicator,
stormy/mixed/fair to generate two instruments. This leads to TSLS
estimates equal to
\[
\hat\beta_{1,\tsls} = -1.014\quad  (\mbox{s.e. }0.384).
\]

%sA.6 #&#
\subsection{Limited-Information-Maximum-Likelihood}

The second most popular estimator in this over-identified setting is
the limited-information-maximum-likelihood (LIML) estimator, originally
proposed by Anderson and Rubin (\citeyear{A49}) in the statistics literature.
The likelihood is based on joint normality of the joint endogenous
variables, $(Y_i,X_i)'$, given the instruments and exogenous variables
$(Z_i,V_i)$:
\[
\left( %
\begin{array} {@{}c@{}} Y_i
\\
X_i \end{array} %
 \right) \bigg| Z_i,V_i
\sim\mathcal{N} \left( \left( %
\begin{array} {@{}c@{}} \vspace*{2pt}\pi_{10}+
\beta_1\pi_{21}'Z_i+
\pi_{12}'V_i
\\
\pi_{20}+\pi_{21}'Z_i+
\pi_{22}'V_i \end{array} %
 \right),
\Omega \right).
\]
The LIML estimator can be expressed in terms of some eigenvalue
calculations, so that it is computationally fairly simple, though more
complicated than the TSLS estimator which only requires matrix inversion.
Although motivated by a normal-distribution-based likelihood function,
the LIML estimator is consistent under much weaker conditions, as long
as $(\varepsilon_{1i},\varepsilon_{2i})'$ are independent of
$(Z_i,V_i)$ and the model (\ref{eq:outcome}) is correct with
$\varepsilon_i$ independent of $(Z_i,V_i)$.

Both the TSLS and LIML estimators are consistent and asymptotically
normally distributed with the same variance. In the just-identified
case, the two estimators are numerically identical. The variance can be
estimated as in the just-identified case as
\[
\widehat{\mathbb{V}}=\hat\sigma^2_\varepsilon\cdot \left( \left(
\begin{array} {@{}c@{}} 1
\\
\hat X_i
\\
V_i \end{array} %
 \right) \left( %
\begin{array} {@{}c@{}} 1
\\
\hat X_i
\\
V_i \end{array} %
 \right)' \right)^{-1} .
\]
In practice, there can be substantial differences between the TSLS and
LIML estimators when the instruments are weak (see Section~\ref{section_weak}) or when there are many instruments (see Section~\ref{section_many}), that is, when the degree of overidentification is high.

For the fish data, the LIML estimates are
\[
\hat\beta_{1,\liml} = -1.016\quad  (\mbox{s.e. }0.384).
\]

%sA.7 #&#
\subsection{Testing the Over-Indentifying Restrictions}

The indirect least squares procedure does not work well in the case
with multiple instruments. The two reduced form regressions are
\[
X_i=\pi_{20}+\pi_{21}'Z_i+
\pi_{22}'V_i+\varepsilon_{2i}
\]
and
\[
Y_i=\pi_{10}+\pi_{11}'Z_i+
\pi_{12}'V_i+\varepsilon_{1i}.
\]
If the model is correctly specified, the $K$-component vector $\pi
_{11}$ should be equal to $\beta_1\cdot\pi_{21}$. However, there is
nothing in the reduced form estimates that imposes proportionality of
the estimates. In principle, we can use any element of the
$K$-component vector or ratios $\hat\pi_{21}/\pi_{11}$ as an
estimator for $\beta_1$.
If the assumption that $\varepsilon_{1i}$ is independent of $Z_i$ is
true for each component of the instrument, all estimators will estimate
the same object, and differences between them should be due to sampling
variation.
Comparisons of these $K$ estimators can therefore be used to test the
assumptions that all instruments are valid.

Although such tests have been popular in the econometrics literature,
they are also sensitive to the other maintained assumptions in the
model, notably linearity in the endogenous regressor and the constant
effect assumption. In the local-average-treatment-effect set up from
Section~\ref{section_late}, differences in estimators based on
different instruments can simply be due to the fact that the different
instruments correspond to different populations of compliers.

\end{appendix}


\section*{Acknowledgments}
Financial support for this research was generously provided
through NSF Grants 0820361 and 0961707. I~am grateful to Joshua
Angrist who got me interested in these topics many years ago, and over
the  years, has taught me much about the issues discussed in this
manuscript, the editor of \textit{Statistical Science} for suggesting this
review and three anonymous referees who wrote remarkably thoughtful
reviews.

%suskaldyti doi

%
%

% imsref loaded by daiva.urboniene, 2014-07-24 14:08:52
% imsref loaded by daiva.urboniene, 2014-07-25 09:10:23
%
% imsref loaded by daiva.urboniene, 2014-07-30 08:59:14

%
\end{document}
